\documentclass{aa} 
\usepackage{pdflscape} 
\usepackage{float}
\usepackage{natbib}
\usepackage{graphicx}
\usepackage{dblfloatfix}
\usepackage{booktabs}
\usepackage{rotating} 
\usepackage{pdflscape} 
\usepackage[toc,page]{appendix}
\usepackage{txfonts}
\usepackage{subcaption}
\usepackage[font=footnotesize]{caption}
\usepackage[labelfont=bf]{caption}
\usepackage[colorlinks=true, citecolor=blue, urlcolor=red, linkcolor=black]{hyperref}

\begin{document} 
\title{Protostellar evolution of intermediate-mass stars: \\ implications for the survival of fossil magnetic fields}
\titlerunning{Protostellar evolution of intermediate-mass stars}
\author{Luis Villarroel Casanova
       \inst{1}, 
       Dominik R.G. Schleicher\inst{2}, 
       Daniele Galli\inst{3} \&
       Jaime A. Rosales\inst{1}}

\authorrunning{Villarroel et al.}
\institute{Departamento de Astronom\'ia, Facultad Ciencias F\'isicas y Matem\'aticas,    Universidad de Concepci\'on, Av. Esteban Iturra s/n Barrio Universitario, Casilla 160-C, Concepci\'on, Chile\\
\email{lvillarroel2020@udec.cl}
\and
Dipartimento di Fisica, Sapienza Universit\`a di Roma, Piazzale Aldo Moro 5, 00185 Rome, Italy\\
\email{dominik.schleicher@uniroma1.it}
\and
INAF Osservatorio Astrofisico di Arcetri, Largo E. Fermi 5, 50125 Firenze, Italy}

\date{Received December 3, 2025; accepted October 6, 2026}

 \abstract
{The {detection} of strong magnetic fields in radiative stars, particularly {in} chemically peculiar Ap/Bp-type stars, has led to the development of different {theoretical scenarios} about their origin. Among them, the fossil field theory offers a possible explanation. However, the survival of such fields is intrinsically dependent on the stellar structure. Studying the transition from a convective to a radiative stage on the stellar structure is extremely relevant to understand whether {a fossil magnetic field} can be preserved throughout protostellar evolution.}
{Our {aim} is to {determine} the {characteristic} mass of protostars {at} which the transition from a convective to a radiative state occurs, as the emergence of a predominantly radiative structure {during accretion may favor the survival of magnetic fields}.}
{We study the pre-main-sequence evolution of intermediate-mass protostars, {starting from convective initial models}, using the Modules for Experiments in Stellar Astrophysics, a 1D stellar evolution model, investigating the dependence on metallicity and average accretion rate, considering different scenarios for the accretion history. }
{We find that the transition mass from a predominantly convective to a radiative state depends {to some extent} on both the average accretion rate and protostellar metallicity, occurring between $\sim1.3\,M_\odot$ and $2.8\,M_\odot$ {for our reference models with an initial mass of $1\,M_\odot$ (up to $\sim4\,M_\odot$ for other initial masses at the highest accretion rates)}. {For the adopted polytropic initial models and the accretion-rate histories imposed after initialization,} the details of the accretion history have a relatively limited impact on the transition. At lower metallicity, the transition occurs at a lower mass. For metallicities $Z\geq0.01$ and accretion rates above $4\cdot10^{-6}\,M_\odot\,\mathrm{yr}^{-1}$, several models retain a convective outer envelope up to {the stopping point of the calculation}, {in which case} any accreted magnetic fields might be subject to dissipation.}
{Since the transition usually occurs between $1.3\,M_\odot$ and $2.8\,M_\odot$ {for the reference models}, our results, within the adopted polytropic framework, provide a possible explanation for the preferential survival of magnetic fields in more massive stars, consistent with observations, {provided that the radiative structure persists after accretion ceases, a phase we do not follow here}. A further caveat is that, in models with high accretion rates, any magnetic field advected with the accretion flow may still be dissipated within the convective outer layers. An additional testable prediction is that the fraction of radiative stars with strong magnetic fields should increase at metallicities below $Z=0.01$.}

\keywords{Stars: magnetic fields, stars: formation, stars: chemically peculiar}

\maketitle

\section{Introduction}
Despite the ubiquity of stars, their formation and evolution {remain fundamental} challenges in astrophysics. Among the many open questions, the origin of their magnetic fields {remains particularly elusive}. In convective and rotating stars, magnetic fields can generally be explained by the presence of a dynamo mechanism \citep{Brandenburg2005, Kapyla2023}, producing dynamo cycles and potentially complex magnetic field structures. 

Strong magnetic fields have been observed over the years in chemically peculiar intermediate-mass stars during their main-sequence (hereafter MS) phase, which make up a small fraction of the A-type star population. These radiative stars, often classified as Ap/Bp stars, have masses ranging from 1.5 to 6 ${M}_{\odot}$ and exhibit large-scale magnetic fields with average strengths from 200 G to 30 kG \citep{Auriere2007}. Their origin so far is not yet well understood. In relation to this, there is also a classification for chemically peculiar stars from the upper main sequence, called CP stars \citep{Preston1974}. Since the pioneering calculations made by \citet{Cowling1945}, is known that large-scale magnetic fields could persist in such stars for considerably longer than their MS lifetimes. This longevity suggests a possible primordial origin, consistent with the hypothesis of fossil magnetic fields \citep[e.g.][]{Braithwaite2017, Schleicher2023}. In this case, the magnetic fields must have survived the pre-main-sequence phase of stellar evolution, which at least preferentially is more feasible when the protostars are radiative as well. Here in the following, we shall first discuss the general properties of the Ap/Bp-type stellar population, before considering the possible origin of their magnetic fields and the complexity of protostellar evolution.

 
 Observations in a volume-limited sample within 100\,pc from the Sun have shown that the fraction of magnetic Ap/Bp stars increases with mass: it is below 1\% at \(1.5\,{M}_{\odot}\) and rises to \(\geq 20\%\) at \(3.5\,{M}_{\odot}\). Below \(1.5\,{M}_{\odot}\) (spectral type F0), strong magnetic fields are absent, probably due to the presence of efficient convective envelopes \citep{Power2007}. Other properties also correlate with the presence of very strong magnetic fields. As shown by \citet{Mathys2008}, Ap stars, in particular, exhibit an astonishing range of rotation periods spanning at least four orders of magnitude, and these fields are found only in stars with rotation periods shorter than 150 days. Therefore, extremely strong magnetic fields, in excess of 7.5\,kG, do not occur in very slowly rotating stars, with periods longer than 150 days. 
Interestingly, according to \citet{Braithwaite2017}, the chemical peculiarities seen in these stars appear to develop after the magnetic field is already in place, likely during a stage in the pre-main-sequence  (hereafter PMS) phase. This suggests that the PMS phase may play a critical role in setting up the magnetic field and the chemical characteristics that we observe. 

In the PMS phase, protostars are classified as T Tauri stars if they are later than spectral type F5 and as Herbig Ae/Be (HAeBe) stars if they are earlier \citep{Braithwaite2017}. These types of stars are typically found in regions of ongoing or recent star formation. Their position on the Hertzsprung-Russell (HR) diagram is close to the MS. Their spectra exhibit emission lines, commonly interpreted as originating from accretion disks \citep{Donati2009}. Essentially, they are precursors of MS A- and B-type stars. After a certain period, their interiors become fully radiative, meaning they no longer possess the convective layers required to sustain a dynamo mechanism, unlike their lower-mass, convective counterparts, the T Tauri stars. In a study published by \citet{Alecian2013}, which examined 70 HAeBe stars, it became clear that a fraction of these stars exhibit magnetic fields. This suggests a different origin for these fields, as traditional dynamo action is not expected in fully radiative stars.


According to the non-ideal magnetohydrodynamics (MHD) equations, the magnetic field \textbf{B} evolves as
\begin{eqnarray}\label{MHD}
    \frac{\partial {\textbf{B}}}{\partial t}&=&\nabla\times (\textbf{u}\times{\textbf{B}}-\eta\nabla\times \mathbf{B}), 
\end{eqnarray}
where $\textbf{u}$ denotes the fluid velocity, $\textbf{B}$ the magnetic field, and $\eta$ the reciprocal of the electrical conductivity, the magnetic diffusivity. The fluid velocity evolves according to the momentum equation:
\begin{eqnarray}\label{field}
    \frac{d{\textbf{u}}}{dt}&=&-\frac{1}{\rho}\nabla P + \mathbf{g} + \frac{1}{4\pi \rho}(\nabla\times \mathbf{B})\times \textbf{B},
\end{eqnarray}
where $d/dt$ {denotes the material derivative, following the motion of the fluid in the Lagrangian frame}, $\rho$ is the density, $\nabla P$ is the pressure gradient, $\textbf{g}$ the gravitational acceleration, and the last term of Eq. \eqref{field} corresponds to the Lorentz force. 

In an unmagnetized star in hydrostatic equilibrium, the pressure gradient balances gravity, resulting in a stationary configuration. However, for a magnetized star, a stable equilibrium will be reached when the pressure gradient, gravity, and the Lorentz force are in balance. In such a case, the system can settle in a static configuration with $|\textbf{u}|=0$ that also needs to be stable, allowing the possible survival of a {fossil magnetic field} \citep{Braithwaite2017}. Bearing that in mind, the induction equation reduces to
\begin{eqnarray}\label{fossileqn}
    \frac{\partial \textbf{B}}{\partial t} &=& \eta\nabla^{2}\textbf{B}.
\end{eqnarray}
The fossil field hypothesis was first proposed by \citet{Cowling1945}, who showed that due to the resistive (or Ohmic) decay time, magnetic fields in radiative stars can survive for durations longer than their MS lifetimes. In the case of the Sun, Cowling realized that the Ohmic diffusion time in radiative zones is around $10^{10}$~years. This decay timescale is approximately given by $\sim R^2/\eta$, where $R$ is the stellar radius and $\eta$ is the magnetic diffusivity. 

Cowling concluded that stellar magnetic fields could be remnants of magnetic fields of the primordial protostellar cloud from which they formed \citep{Cowling1945, Dudorov2015}. Therefore, a magnetic field could survive the whole MS if it is in a stable equilibrium inside a radiative zone. Some of these stable configurations have already been simulated inside fully radiative A-type stars \citep{Braithwaite2004,Braithwaite2006,Braithwaite2008,Becerra2022}. However, the survival of fossil fields is strongly dependent on the internal structure of the star. {In stars that undergo a predominantly or fully convective protostellar phase, turbulent motions can efficiently dissipate a substantial fraction of the initial magnetic flux}. In contrast, if the interior is fully radiative or if the fossil field is strong enough to suppress the convection in the core, it may survive for a much longer time, comparable to the lifetime of the star \citep{Moss1989, Schleicher2023}. 

The transition from a convective to a radiative phase in the protostellar structure is {a crucial factor in determining} the survival of these primordial magnetic fields. {The development of a radiative interior may allow} to preserve a significant part of the initial magnetic flux of the protostellar cloud, as envisioned by \citet{Dudorov1990}. The first detailed protostellar models for the intermediate mass that showed such a transition were proposed by \citet{Palla1990,Palla1991,Palla1992}.

An alternative scenario is provided by the core dynamo theory, which proposes that a dynamo mechanism operates within the stellar convective core and continuously regenerates the magnetic field that then is transported into the overlying radiative layers, {driven by} buoyancy, towards the surface \citep{Braithwaite2017}. Specifically, in early-type stars, the formation of this convective core is driven by the temperature sensitivity to the CNO cycle. Under these conditions, convective motions and differential rotation can produce shear instabilities that can potentially generate strong magnetic fields through dynamo processes \citep{Browning2004, Brun2005}. According to estimations made by \citet{Lecoanet2022}, a convective core dynamo could generate strong magnetic fields inside a B-type star, with an upper limit of $B_r \approx 500 \ \mathrm{kG}$ to its strength, exceeding what is possible to reach via fossil fields alone. Nevertheless, the {operation} of such a dynamo does not guarantee that the magnetic field will reach the stellar surface, forming coherent large-scale structures \citep{MacDonald2004}, as only a fraction of the magnetic flux is {expected to be} effectively transported outward \citep{MacGregor2003,Hidalgo2024}. There is also the possibility of interaction between this primordial field and the one generated by the core dynamo. In fact, \citet{Featherstone2009} modeled the coupling of a fossil field and a convective core dynamo within an A-type star, demonstrating that this interaction could produce a magnetic field strong enough to be transported to the surface by magnetic buoyancy, enhancing the action of the dynamo \citep{MacDonald2004}. The impact of this interaction on the fossil magnetic field, both at the surface and within the radiative envelope, was recently investigated by \citet{Hidalgo2025}, who found that certain configurations can enhance the core dynamo and lead to stable surface fields.


We know that stars form through the fragmentation and subsequent gravitational collapse of cold, dense, and massive molecular clouds. During this process and because the collapse has been found to be highly non-homologous \citep{McNally1964, Bodenheimer1968, Larson1969}, a small central core reaches hydrostatic equilibrium, while the rest of the cloud continues to collapse. The remaining material is accreted onto the central core. This process essentially describes the formation of a pre-main-sequence (PMS) star in a scenario driven by accretion \citep{Stahler1980a, Palla1990, Palla1991, Palla1992}.

 Since the breakthrough period in the mid-20th century, marked by foundational contributions to stellar structure and nucleosynthesis \citep{Burbidge1957, Bohm-Vitense1958, Schwarzschild1958, Henyey1964, Paczynski1970A}, as well as pioneering studies on the evolution of PMS stars \citep[e.g.,][]{Larson1969, Baraffe1998, Palla2000}, the field of protostellar evolution has seen tremendous progress. In particular, numerous stellar evolution codes have been developed to address a broad range of astrophysical problems, enabling detailed modeling across a wide range of stellar masses and evolutionary phases. Notable examples include the Geneva code \citep{Ekström2012, Georgy2013}, STARS \citep{Eggleton1971, Pols1995}, STERN \citep{Brott2011}, and the Padova/PARSEC \citep{Girardi2002, Bressan2012} models. The increasing availability of precise tabulated opacities, nuclear reaction rates, and equations of state has facilitated major advances in the study of stellar astrophysics \citep{Choi2016}. 

 Observational {studies} have established a correlation between the magnetic field strength $ B$ and the {gas number} density $ n$, for star-forming clouds. {This relationship can be expressed as}
 \begin{equation}\label{mag}
     \frac{B}{B_0} = \left(\frac{n}{n_0}\right)^{k}.
 \end{equation}
 Here, $B_0$ and $n_0$ depend on the cloud formation conditions. For a density range between $1$ and $10^{10} \ \text{cm}^{-3}$, a large part of the observational data can be described with the following parameters,\begin{equation}
     B_0=5 \mu \mathrm{G}, n_0=50-200 \ \text{cm}^{-3}, k = 1/2 -2/3.
 \end{equation}
This is {supported} by optical and infrared {polarimetric observations} of dense molecular and protostellar clouds \citep{Vallee1997, Girart2006, Li2009, Crutcher2012, Chapman2013, Koley2022}. A comparison of the observed magnetic flux for radiative stars and interstellar molecular clouds suggests that star formation occurs in magnetized clouds. Hence, the magnetic flux of young and some MS stars may be a relic of molecular or protostellar clouds, implying a fossil origin \citep{Dudorov2015}. Several numerical simulations have been carried out to study the evolution of the magnetic field strength during gravitational collapse, with initial strengths of $30 - 300 \ \rm \mu G$ at a number density of $300 \ \text{cm}^{-3}$. When molecular clouds collapse, the magnetic field strength increases as the density of the gas rises, leading to a central field strength of about $0.1$~G at a number density in the range of $10^{10}-10^{11}\ \text{cm}^{-3}$ \citep{Desch2001, Nakano2002, Tomida2015, Masson2016}. However, {at sufficiently high densities,} the field stops increasing and is no longer efficiently coupled to the gas, and remains approximately constant while the gas increases towards protostellar densities of $10^{15}-10^{16} \  \text{cm}^{-3}$, due to a drop in the degree of ionization and a decoupling of the magnetic field from the gas. Only when the temperature rises sufficiently to ionize the gas does the magnetic field couples again to the gas  \citep{Nakano1988,Nakano_ubemayashi1988}. The fields that are amplified in this way may be accreted onto protostars and may become the fossil fields observed in Ap/Bp-type stars.

In this work, we aim to determine when and under what conditions protostars go through a relevant radiative phase, which considerably increases the chance of survival of the magnetic fields. This is particularly relevant as accretion onto the protostars is typically a continuous process and very likely to proceed even after this transition. Here, we explore protostellar evolution considering different scenarios for the accretion rate, including constant, decaying, and oscillating accretion rates. We computed the protostellar evolution using Modules for Experiments in Stellar Astrophysics (MESA; \cite{Paxton2011, Paxton2013, Paxton2015, Paxton2019, Jermyn2023}), a widely used 1D stellar evolution code that is particularly well suited for this purpose. We use the simulations to {determine the protostellar mass at which} the transition occurs from a convective to a radiative state, which we explore as a function of the mean accretion rate and {considering} different {prescriptions for} the time dependence of the accretion rate and the metallicity.

The remainder of {this} paper is organized as follows. In Section \ref{Method}, we provide an overview of the MESA code, focusing on the control parameters and describing our physical model for the evolution of the PMS with different types of accretion prescriptions, including constant and time-dependent rates. In Section \ref{results}, we examine the influence of the accretion rate on the stellar interior. Our conclusions and discussions are summarized in Section \ref{discusionandconclusion}.

\section{Method}\label{Method}
To simulate the protostellar evolution, we employ the MESA stellar evolution code to solve the fundamental stellar structure equations. {In this Section}, we describe the modeling and the criteria used to analyze the transition from the convective to the radiative stage during this section, as well as the models we consider for the time-dependent accretion. As a final point, we introduce the concept of the transition mass, which we define as the protostellar mass at which the transition occurs from a predominantly convective to a predominantly radiative state.

\subsection{The MESA code for stellar evolution}
To model the evolution of protostars, we use the open-source 1D stellar evolution code MESA version r23.05.1 \citep{Paxton2011,Paxton2013,Paxton2015,Paxton2019, Jermyn2023}, which calculates models for spherically symmetric single stars. Its 1D stellar evolution module, MESAstar, employs a Newton-Raphson solver to simultaneously solve the fully coupled Lagrangian equations governing stellar structure and composition, where the independent variable is the mass $M_r$ enclosed within radius $r$. 

\begin{eqnarray}\label{continuity}
    \frac{\partial r}{\partial M_r} &=& \frac{1}{4 \pi r^2 \rho}, \\ \label{momentum}
    \frac{\partial P}{\partial M_r} &=& -\frac{G M_r}{4 \pi r^4}, \\ \label{energy}
    \frac{\partial L_{r}}{\partial M_r} &=& \epsilon_{\text{nuc}} - \epsilon_\nu - T \frac{\partial s}{\partial t}, \\ \label{temperature}
    \frac{\partial T}{\partial M_r} &=& -\frac{G M_r T}{4 \pi r^4 P} \nabla,\\ \label{composition}
    \frac{\partial X_i}{\partial t} &=&\frac{m_i}{\rho} \left(\sum_{j} r_{ji} -\sum_{k} r_{ik} \right) \text{[ + mixing]}, \ i = 1, \ldots, N.
\end{eqnarray}

The stellar structure is described by the continuity \eqref{continuity}, momentum \eqref{momentum}, energy \eqref{energy}, temperature gradient \eqref{temperature}, and composition \eqref{composition} equations. Where, $r$ is the radial coordinate, $\rho$ is the density, $P$ is the pressure, $L_r$ is the luminosity in radius $r$, $T$ is the temperature, and $s$ is the specific entropy. The term $\epsilon_{\text{nuc}}$ denotes the rate of nuclear energy generation per unit mass and $\epsilon_\nu$ corresponds to the energy loss rate due to neutrinos. The symbol $\nabla$ represents the temperature gradient, while $X_i$ is the mass fraction of the $i$-th chemical species, with $m_i$ its atomic mass. The reaction rates are indicated by $r_{ji}$ and $r_{ik}$, and the term ``[+ mixing]'' indicates that chemical mixing processes may be included, depending on the physical model. The code employs mixing length theory (MLT) to determine the temperature gradient and the Ledoux criterion for convection.

In this study, we calculate models of protostellar evolution with an initial mass of one solar mass for the main set of models (other initial masses are explored in Appendices~\ref{apendixg} and~\ref{apendixh}), and different metallicities set at $Z= 0.001, 0.01, 0.02$ and $0.03$. In the main model grid, the initial deuterium abundance is set to zero so that deuterium burning is not included. Appendix~\ref{apendixh} shows that including deuterium has only a weak effect on the transition mass. We use the fixed mixing length parameter $\alpha_{\text{MLT}} = 1.82$ from \citet{Choi2016}, which is close to the solar value, and parameterize the efficiency of convection inside the star. We adopt the modified version of MLT from \citet{Henyey1965}, which takes into account radiative losses from fluid elements. We adopt an overshooting parameter \( f_{\mathrm{ov}} = 0.016 \), following \citet{Choi2016}, which sets the extent of the overshoot region beyond the convective boundary defined in Eq. (9) of \citet{Paxton2013}. Although its impact may be negligible in the early stages of stellar evolution, it can become relevant in later stages when there is a convective core. Finally, we adopt the default MESA {prescriptions} for the equation of state and the opacities. The equation of state is mainly based on tabulated data from OPAL and SCVH {databases} \citep{Rogers2002, Saumon1995}. For radiative opacities, we use the high-temperature tables based on \citet{Grevesse1998} and the low-temperature tables from \citet{Ferguson2005}. The electron conduction opacities are calculated from the prescription of \citet{Blouin2020}. The opacity tables adopted in this study differ from those used in some of the earlier studies in this field \citep[e.g.][]{Palla1991, Palla1992}, which may result in differences in the predicted evolutionary properties.

\subsection{Modeling the accretion onto the protostar} \label{modeling}
MESA has a unique feature among stellar evolution codes, which is the advantage of having a modular structure. This allows us to experiment with the available physics in the simulation and to implement new physical prescriptions in a straightforward way. For the simulations pursued here, we have implemented a module for a time-dependent accretion rate, which is defined via the desired functional form and including the possibility of the rate being constant.

We assume that the composition {of the accreted material} is identical to that of the current stellar surface and, consequently, that the entropy of the accreted material matches the entropy of the stellar surface prior to accretion. This corresponds to the assumption of cold disk accretion \citep{Palla1992,Hosokawa2010,Haemmerle2013}, which means that matter is gently deposited onto the stellar surface. In this way, we will consider constant rates as well as time-dependent accretion rates, including a power-law decrease and time-dependent oscillations, as outlined below.

All models studied here are evolved until they reach a stopping condition, defined using the MESA control parameter \texttt{stop\_near\_zams}, which is triggered when $L_{\rm nuc}/L>0.9$. {We refer to this endpoint as the stopping point of the calculation. In models with high accretion rates, $L_{\rm acc}$ can remain comparable to or larger than the stellar luminosity, so this condition does not guarantee that the classical ZAMS is reached (see Sect.~3.4).} This endpoint is adopted for clarity; however, for the purposes of this study, it is not essential to {establish} whether the ZAMS is actually reached, as our primary objective is to identify the transition point at which the protostar enters the radiative phase.

Regions that appear convectively unstable under the Schwarzschild criterion (hereafter {SC}) can actually be stable when composition gradients are considered. This condition defines what it is known as semiconvection, which is included in the code by specifying the use of the Ledoux criterion to treat convection in simulations. However, for this project, the {SC} is sufficient to quantify the transition between radiative and convective phases, since composition gradients are not significant.

To analyze the internal structure of the protostars, we use the Kippenhahn diagrams, similar to \citet{Palla1991,Palla1992}, also called space-time diagrams, to illustrate the evolution of the protostellar interior and to determine the convective and radiative regions. 
The main equation that MESA has to solve in order to obtain the necessary data is Eq. \eqref{temperature}. The gradient {is given by} the following conditions:
\begin{eqnarray}\label{eqn11}
     \nabla = 
    \begin{cases} 
        \nabla_{\text{rad}} = \frac{3 \kappa}{16 \pi a c G} \frac{L_r P}{M_r T^4}, & \text{if } \nabla_{\text{rad}} \leq \nabla_{\text{ad}} \\ 
        \nabla_{\text{ad}} + \Delta \nabla, & \text{if } \nabla_{\text{rad}} > \nabla_{\text{ad}}
    \end{cases} 
\end{eqnarray}
 where $\nabla_{\rm rad}$ and $\nabla_{\rm ad}$ are the radiative and adiabatic temperature gradients, respectively, and $\Delta \nabla$ refers to the superadiabatic excess\footnotemark[1] in a convective region. {Here}, $\kappa$, $a$, $G$ and $c$ denote the opacity, the radiation constant, the gravitational constant and the speed of light, respectively. Eq. \eqref{eqn11} defines the temperature gradient $\nabla$ inside the star and takes two possible forms depending on whether the radiative gradient $\nabla_{\text{rad}}$ is less than or equal to the adiabatic gradient $\nabla_{\text{ad}}$. If $\nabla_{\text{rad}} < \nabla_{\text{ad}}$, the energy is transported by radiation and $\nabla$ equals $\nabla_{\text{rad}}$, which is given by the expression above. If $\nabla_{\text{rad}} > \nabla_{\text{ad}}$, convection sets in and the gradient is $\nabla_{\text{ad}} + \Delta \nabla$. Therefore, on the basis of the temperature gradients and applying the {SC}, we can distinguish between radiative and convective regions within the protostar. The quantity $\nabla_{\text{rad}} - \nabla_{\text{ad}}$ is known as superadiabaticity, {and} allows us to determine the radiative fraction of the protostar. We define the structural transition point as the stage when 50\% of the stellar interior becomes radiative.
We further define the transition mass $M_{\text{tr}}$ as the mass that a protostar reaches at this transition point. 

{In} this work, we will determine how the transition mass for this transition depends on metallicity as well as the time-averaged accretion rate during protostellar evolution, considering scenarios with time-dependent, {oscillating}, and decreasing accretion rates. 

\footnotetext[1]{This is the correction applied when convection is not perfectly adiabatic.}

\section{Results}\label{results}
In this section, we present the results {of our analysis of} protostellar evolution {under} different scenarios, {considering both} to constant and time-dependent accretion rates. Our main {goal} is to derive the transition mass for which at least half of the protostar becomes radiative. We assess here how it depends on the average accretion rate, on metallicity, and on the {adopted accretion history}. We {initialize the calculation} with a protostar of $1 M_{\odot}$, in hydrostatic {equilibrium} with a central temperature of $10^5$K, thereby ensuring {that} nuclear reactions are negligible. The internal structure is derived from an $n=1.5$ polytrope, {yielding} a chemically homogeneous and fully convective {configuration} characterized by a flat (adiabatic) entropy profile. We also checked that our results {depend only weakly on the initial mass assumed for the evolution at accretion rates $\lesssim 10^{-5}\,M_\odot\,\mathrm{yr}^{-1}$, while the dependence becomes stronger at higher accretion rates (see Fig.~\ref{fig:lowerr})}.

\subsection{Constant accretion rates}\label{constantrate}

Our simplest scenario for accretion corresponds to a constant accretion rate $\dot{M}$ on the protostar. In Figure \ref{kipp} we show four examples for different cases of a constant accretion rate. The initial protostellar mass corresponds to $1$$M_\odot$ and it is assumed to have solar metallicity (Z=0.02). We explore accretion rates ranging from $10^{-6}$ to $1.5\cdot 10^{-4}$ $M_\odot$~yr$^{-1}$.  In {the Kippenhahn} diagram, the horizontal axis corresponds mass evolution and the vertical axis to the radial coordinate inside the star. The green-hatched area indicates the {convective} protostellar region, while the white region {is the radiative zone}. The blue color indicates the energy production within the protostar. The black solid line corresponds to the protostellar radius.

\begin{figure}[h]\label{kipp}
  \centering
  \begin{subfigure}{0.48\columnwidth}
    \centering
    \includegraphics[width=\linewidth]{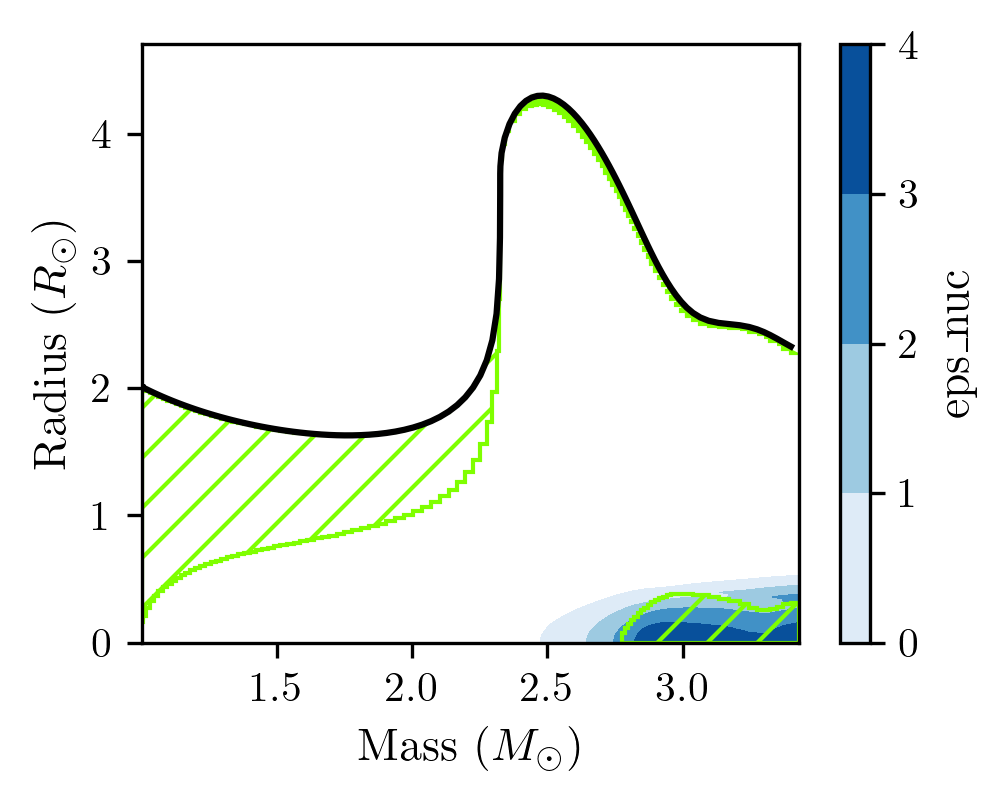}
     \caption{}
     \label{a}
  \end{subfigure}
  \hfill
  \begin{subfigure}{0.48\columnwidth}
    \centering
    \includegraphics[width=\linewidth]{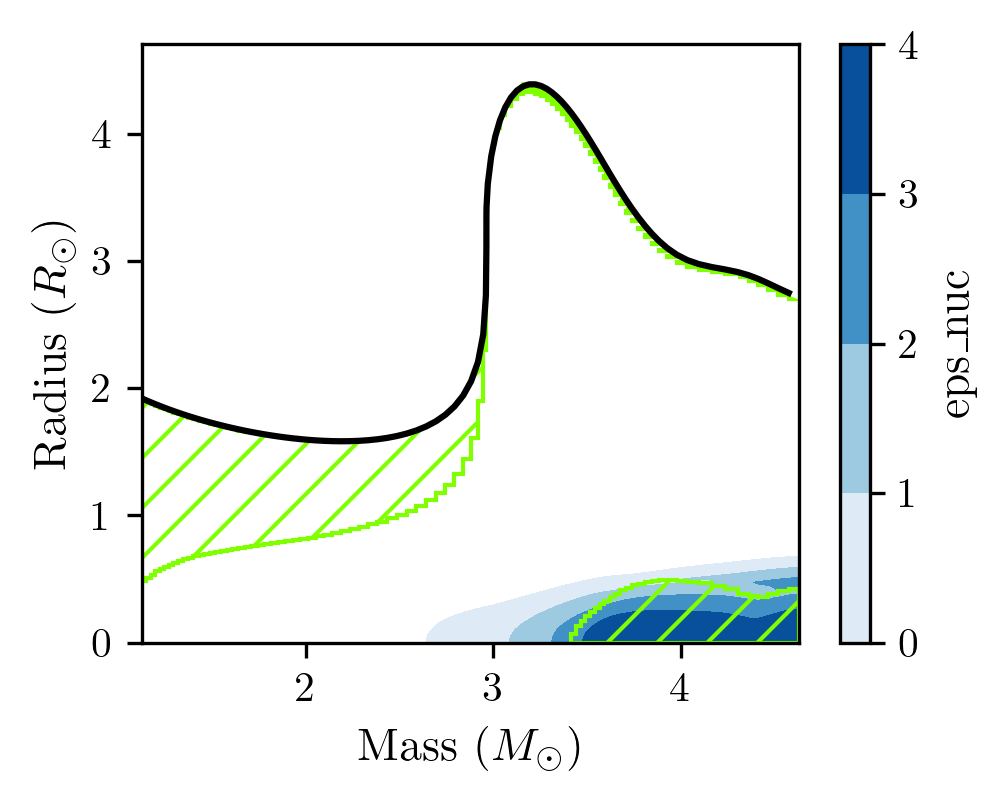} 
    \caption{}
  \end{subfigure}
   \begin{subfigure}{0.48\columnwidth}
    \centering
    \includegraphics[width=\linewidth]{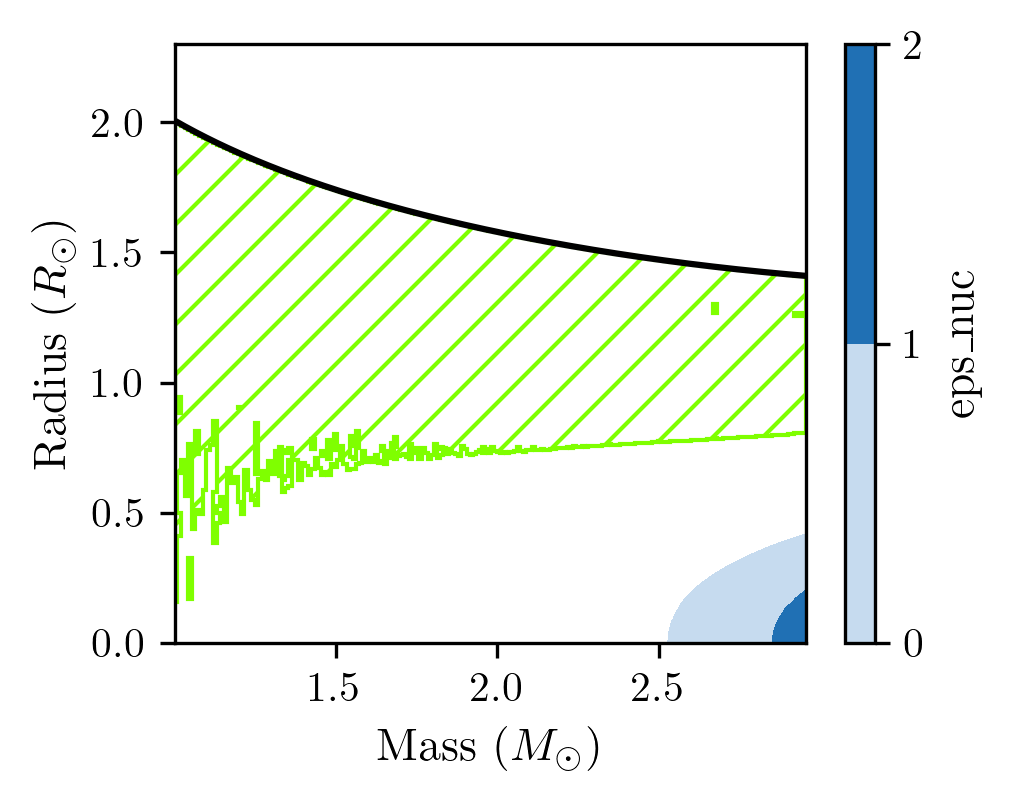}
     \caption{}
  \end{subfigure}
  \hfill
  \begin{subfigure}{0.48\columnwidth}
    \centering
    \includegraphics[width=\linewidth]{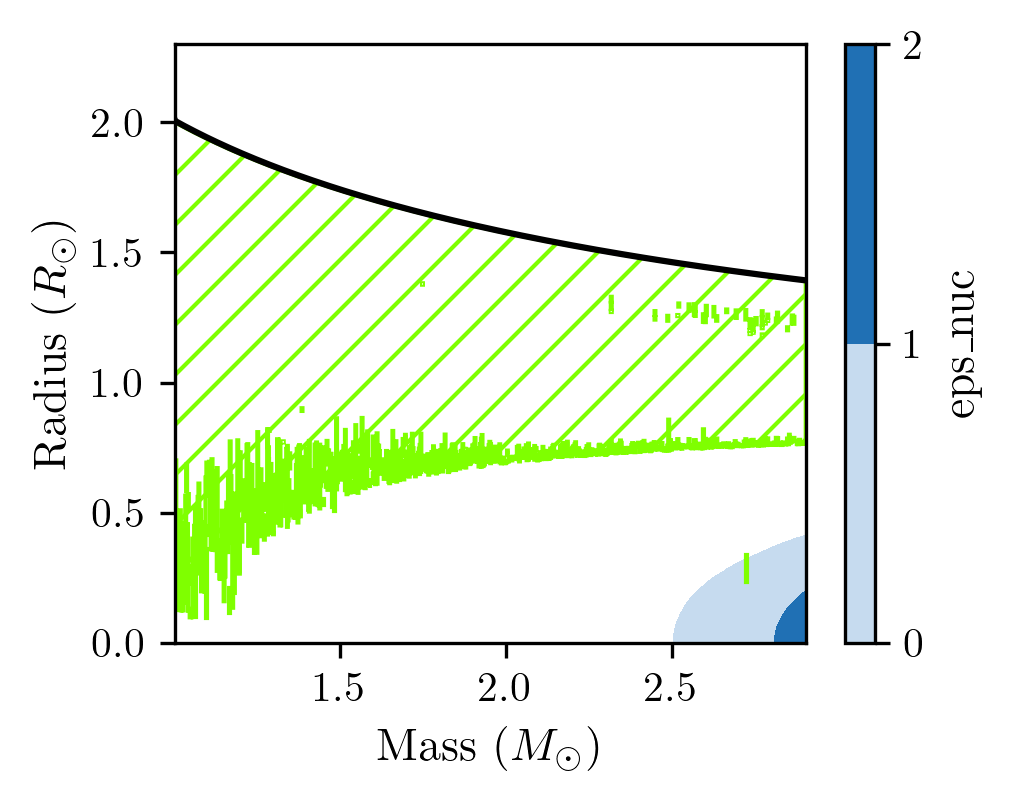} 
    \caption{}
    \label{d}
  \end{subfigure}
  
  \caption{Kippenhahn diagrams for four constant accretion rates at solar metallicity ($Z=0.02$): (a) $\dot{M}=1.00\times10^{-6}$, (b) $3.98\times10^{-6}$, (c) $3.98\times10^{-5}$, and (d) $1.50\times10^{-4},M_\odot,\mathrm{yr}^{-1}$. We show four different models that assume different values of the constant accretion rate $\dot{M}$. The $y$-axis corresponds to the radial coordinate in the protostar, the $x$-axis denotes the mass of the protostar. 
  Green indicates the convective parts of the star, {white} space indicates the radiative zones. The color bar represents the energy generation rate in the central core. All the models are evolved until the stopping condition is reached. {For the lower accretion rates, the stopping point coincides with the ZAMS phase, since $L_{\rm nuc} >L_{\rm acc}$} in those models.}
  \label{kipp}
\end{figure}

The top left panel in Figure \ref{kipp} corresponds to the evolution with an accretion rate of $10^{-6}$$M_\odot$~yr$^{-1}$. In this case, the protostellar radius, initially around $2$~R$_\odot$, increases steeply at $1.5$ million years when energy production begins in the central core and reaches a final value of nearly $3$~R$_\odot$ approaching the zero-age main sequence. At the beginning of the evolution, the protostar is mostly convective except for a small inner region; however, the radiative zone expands and the protostar is fully radiative after about $1.3$ million years. The evolution is qualitatively similar in the top right panel corresponding to an accretion rate of $3.98\cdot10^{-6}$$M_\odot$~yr$^{-1}$. {However}, the corresponding timescales are shorter, as the protostellar radius rapidly expands at about $0.5$ million years, which is also the time when it becomes fully radiative. In the case of a higher accretion rate ($3.98\cdot10^{-5}$$M_\odot$~yr$^{-1}$) (bottom left panel), the behavior changes qualitatively as the radius initially increases as a result of accretion and it decreases steadily as a function of time as the accretion timescale $M/\dot{M}$ becomes longer compared to the Kelvin-Helmholtz timescale for contraction. In this case, the radiative zone of the protostar expands gradually over time, but it never becomes fully radiative. Even {at the stopping point of the calculation}, only about half of the star is radiative. The behavior is similar, but still more extreme for an accretion rate of $1.5\cdot10^{-4}$$M_\odot$~yr$^{-1}$ (bottom right panel). The Kippenhahn diagrams can be employed for analyzing where the transition to the radiative phase occurs (although for a more precise determination, we employed the superadiabaticity as discussed in Section \ref{modeling}). If the transition is found to occur at a time $t_{\rm tr}$, the protostellar mass at that time is given as $M_{\rm tr}=1$~$M_\odot+\dot{M}t_{\rm tr}$, where $\dot{M}$ is the accretion rate of the model. In this way, in Figures \ref{a} to \ref{d}, the different accretion rates correspond to the transition masses $M_{tr}$ of (a) $1.64 \ {M}_{\odot}$,  (b) $ 1.97 \ {M}_{\odot}$, (c) $2.44 \ {M}_{\odot}$, and (d) $2.53 \ {M}_{\odot}$.
 We note that the transition mass slightly increases with the accretion rate, consistent with the results of \citet{Palla1991, Palla1992}.

\begin{figure}[h]
    \includegraphics[width=0.9\linewidth]{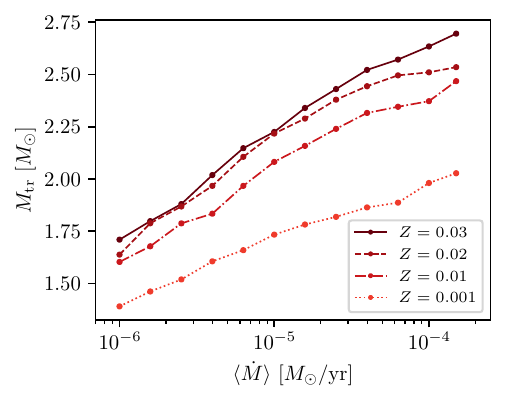}
    \caption{Results for models with constant accretion rate. The transition mass for which at least $50\%$ of the protostar is radiative {is shown} as a function of the average accretion rate as defined via Eq.~\ref{maverage}, equal {here} to the constant accretion rate assumed in the model. The results are given for four different metallicities. }
    \label{constant}
\end{figure}

Our complete suite of simulations with constant accretion rates is presented in Table \ref{tab:summary_constant}. In Figure \ref{constant}, we present the transition mass obtained for 48 different simulations as a function of the accretion rate, for the absolute metallicities of $Z= 0.03, 0.02, 0.01, 0.001$. The shapes of the curves are qualitatively similar but quantitatively different. At first, the transition mass increases steeply with the accretion rate, while it then gradually flattens {increasing} more mildly at later stages. In general, the curves appear to be well described with the following equation: $M_{\mathrm{tr}} =  a \log_{10}\langle \dot{ M} \rangle + b $. The fits to the obtained results are {given} in Table \ref{tab:combined_fits}, where we {list} the parameters $a$ and $b$ for each metallicity. For the high metallicity case ($Z=0.03$), the transition mass starts out at $\sim1.7$$M_\odot$for an accretion rate of 
$\sim 10^{-6}$$M_\odot$~yr$^{-1}$ and increases to $\sim2.7$$M_\odot$ for an accretion rate of $1.5\cdot 10^{-4}$$M_\odot$~yr$^{-1}$. The transition mass decreases with metallicity and for the lowest metallicity case with $Z=0.001$, it starts at $\sim1.4$$M_\odot$ and reaches a final value close to $2$$M_\odot$. 

\subsection{{Oscillating} accretion rates} \label{variable}

We introduce an oscillating accretion rate characterized by periodic variations. These could mimic episodic accretion events or instabilities in the circumstellar disk \citep[e.g.][]{Riaz2018, Riaz2021}. In this case, we assume that the accretion rate consists of a constant component, which we defined as $\dot{M}_0=10^{-6}\  {M}_{\odot}/yr$, plus an oscillating one. We model the oscillating one as a positive contribution, via a square of the sine function, to allow for the peaks of accretion to significantly exceed the constant contribution. For this purpose, we vary the amplitude $A$ and the characteristic period $P$. The time-dependent accretion rate is then given as
\begin{equation}
    \dot{ M}(t)= \dot M_{0} + A\sin^{2}\left(\frac{\pi t}{P}\right).\label{Mfluc}
\end{equation}

For a meaningful comparison with the results obtained with constant accretion rates, we introduce here the time-averaged accretion rate, {defined} as the difference between the initial mass and {the} mass {at which the model stops}, divided by the {time taken to reach the stopping point}:
\begin{equation}\label{maverage}
    \langle \dot M \rangle
=\int_{0}^{t_{\mathrm{stop}}} \frac{\dot M}{t_{\mathrm{stop}}}dt = (M_{\mathrm{stop}} - M_{\mathrm{in}})\cdot\frac{1}{t_{\mathrm{stop}}}.
\end{equation}

\subsubsection{Dependence on metallicity}\label{cteperiod}
\begin{figure}[h]
    \includegraphics[width=0.9\linewidth]{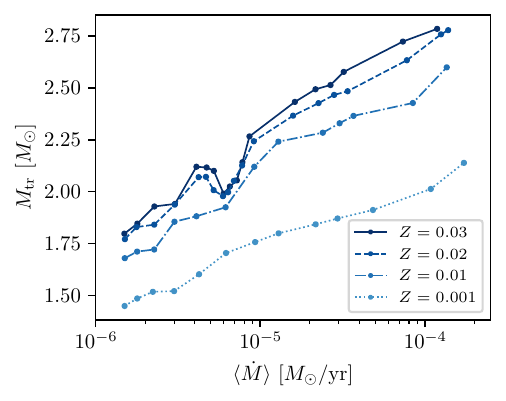}
    \caption{Results for the models with {oscillating} accretion rate as defined in Eq. \eqref{Mfluc}, assuming a fixed period of $50,000$ years. We show here the transition mass for which at least $50\%$ of the protostar is radiative as a function of the average accretion rate defined via Eq.\eqref{maverage}. The results are given for four different metallicities ranging from $0.001$ to $0.03$.}
    \label{fig:difz}
\end{figure}

We have modeled the oscillating accretion rate for a period of $50,000$ years and a range of metallicities from $0.001$ to $0.03$. In Figure \ref{fig:difz}, all models for this setup are shown, and the complete data points are presented in Table \ref{tab:summary_diferent_Z}. In this case, we note that, for lower metallicities, the plot appears to be the same as for the constant accretion rate scenario, starting from $ M_{\text{tr}} = 1.45\ {M}_{\odot}$ at $Z= 0.001$, $ M_{\text{tr}} = 1.68 \ {M}_{\odot}$ at $Z= 0.01$, $ M_{\text{tr}} = 1.77\ {M}_{\odot}$ at $Z= 0.02$ and $ M_{\text{tr}} = 1.80\ {M}_{\odot}$ at $Z= 0.03$ . As the metallicity increases, some differences appear, and between $Z= 0.02$ and $Z= 0.03$ a drop can be seen in the {oscillating} accretion rate scenario, between $\langle  \dot M\rangle = 5 - 9\cdot10^{-6} \ M_\odot$~yr$^{-1}$, in which the transition mass decreases from $2.07$ to $1.98$ solar masses for $Z= 0.02$ and from $2.12$ to $1.99$ solar masses for $Z= 0.03$. Then, {the protostellar mass continues to grow}, reaching a final value of $2.76$ and $2.78$ $M_\odot$, for $Z = 0.02$ and $Z= 0.03$, respectively. However, we note that for the cases with the highest accretion rates, the timescale to reach {the stopping point} becomes comparable to the period of the variations, and consequently, the accretion rate does not vary significantly, This behavior is most evident in the last case.

\subsubsection{Dependence on the timescale of the variations}\label{ctemetallicity}
We explored how the timescale of the variations affects the obtained relation by adopting a constant metallicity $Z= 0.02$, adopting three different periods to explore the influence of these changes on the transition mass of the star. 

\begin{figure}[h]
    \includegraphics[width=0.9\linewidth]{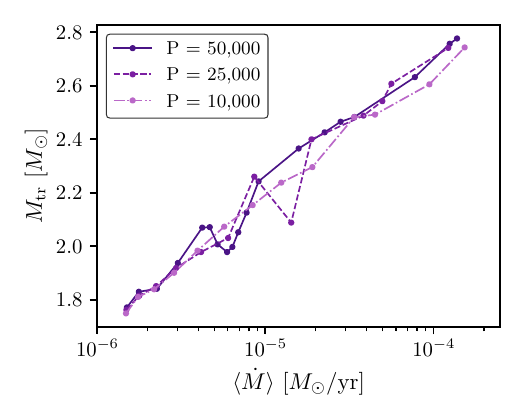}
    \caption{Results for models with {oscillating} accretion rates as defined in Eq. \eqref{Mfluc}, assuming a constant metallicity of $Z=0.02$. We show here the transition mass for the transition to the radiative state as a function of the average accretion rate defined via Eq.\eqref{maverage}. The results are given for three different periods ranging from $10^4$~ years up to $5\cdot10^4$~ years.}
    \label{fig:periods}
\end{figure}

In Figure \ref{fig:periods} the data points from Table \ref{tab:summary_changing_periods} are shown. We note that the plot starts with a very similar transition mass for all three scenarios. We have $ M_{\mathrm{tr}} = 1.77\ {M}_{\odot}$ for $ P=50,000$~yr, $M_{\mathrm{tr}} = 1.76 \ {M}_{\odot}$ for $ P=25,000$ years, and $ M_{\mathrm{tr}} = 1.75\ {M}_{\odot}$ for $P=10,000$ years. As the models evolve, we observe a greater fluctuation between the accretion rates $4\cdot10^{-6}$ - $2\cdot10^{-5}\ M_\odot$~yr$^{-1}$. However, for accretion rates of about $10^{-4} \ M_\odot$~yr$^{-1}$, the transition masses of the respective protostars are again quite similar, with $ M_{\mathrm{tr}} = 2.76 \ {M}_{\odot}$ for $ P=50,000$ years, $ M_{\mathrm{tr}} = 2.61 \ {M}_{\odot}$ for $ P=25,000$ years and $ M_{\mathrm{tr}} = 2.74\ {M}_{\odot}$ for $P=10,000$ years. We therefore conclude that the time period of these variations, at least within the limits explored here, does not affect the relation very significantly. Small dips occur in the cases of periods of $25,000$~years and $50,000$~years, as the transition occurs at a time when the accretion rate is close to a minimum.

\begin{figure}[h]
    \centering
    \includegraphics[width=\linewidth]{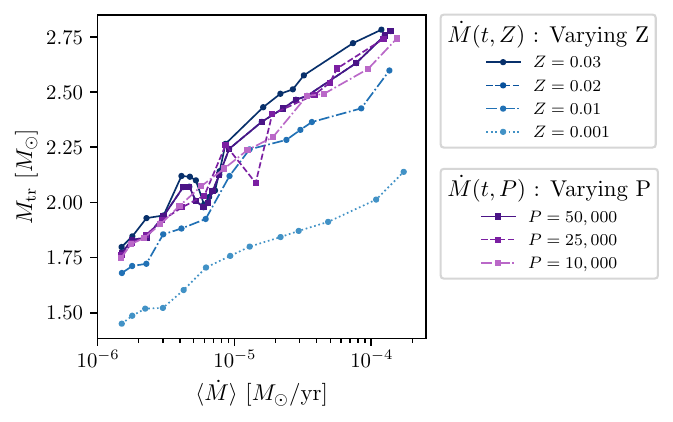}
    \caption{Comparison of models with different metallicities and models with different periods. Blue corresponds to {variations in}  the metallicity, while purple corresponds to{ variations in the oscillation period}.}
    \label{powerlaw:comparison}
\end{figure}

In Figure \ref{powerlaw:comparison}, we compare the results for a constant period and a constant metallicity. We observe that variations in the period of the {oscillation} introduce more changes in the middle part of the plot, but the results remain almost {unchanged} for a given metallicity. In contrast, changes in the metallicity seem to be more relevant, as a clear dependence is observed.

\subsection{Power-law accretion rates}\label{powerlaw}
Finally, we introduce a time-dependent accretion rate following a power-law behavior. This formulation aims to reproduce a scenario in which the accretion rate gradually decreases as the protostar evolves, reflecting the natural decline of material available in the surrounding envelope. Such a model provides a more realistic description of the late stages of protostellar growth compared to the constant or {oscillating} cases. The adopted model is written as
\begin{equation}\label{Mpower}
  \dot{M}=\dot{M}_0\left(\frac{ t+ t_0}{ t_0}\right)^{\alpha}, \quad \text{with}\quad\alpha=-1,-2,-3.
\end{equation}
The values for the initial accretion rate $ \dot{\mathrm{M}}_0$ range from $10^{-4}$ to $10^{-6}\ M_\odot$~yr$^{-1}$. The model starts with an initial mass of $1\ {M}_{\odot}$. Depending on the value of $\alpha$, the characteristic time $ t_0$ corresponds to the moment when the accretion rate drops to half its initial value ($\alpha = -1$), one-quarter ($\alpha = -2$) or one-eighth ($\alpha = -3$). 

To explore the consequences of this prescription, we present again the transition mass as a function of the averaged accretion rate calculated from Eq. \eqref{maverage} for the different models.

\subsubsection{Dependence on metallicity} \label{constant_alpha}

\begin{figure}[h]
    
    \includegraphics[width=0.9\linewidth]{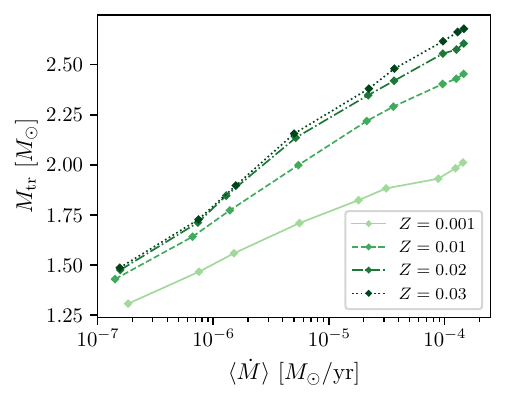}
    \caption{Results for the models with a power-law accretion rate, as defined in Eq. \eqref{Mpower}, assuming $t_0=300,000$~yr and a power-law index of $\alpha = -1$. Here, we show the transition mass for which at least $50\%$ of the protostars are {radiative} as a function of the average accretion rate defined via Eq.\eqref{maverage}. The results are given for four different metallicities.}
    \label{powerlaw:constant_alpha}
\end{figure}

First, we consider models with a fixed parameter $\alpha=-1$ and explore the dependence on metallicity, {varying $Z$ from} $0.001$ to $0.03$. In Figure \ref{powerlaw:constant_alpha}, we show the data point corresponding to Table \ref{tab:summary_powerlaw} for the different metallicities. The models start with a transition mass of $\ M_{\mathrm{tr}}=1.31, \ 1.43, \ 1.48, \ 1.49 \ {M}_{\odot}$ for $Z=0.001, 0.01, 0.02, 0.03$, respectively. In this scenario, for initial accretion rates of the order of $10^{-6}\ M_\odot$~yr$^{-1}$, the average accretion rate reaches a lower value of the order $10^{-7}  \ M_\odot$~yr$^{-1}$. As the average accretion rate increases, the curve behaves very similar to the constant case. For the cases where the metallicity was $Z=0.03$ and $Z=0.02$, the models start with a transition mass very similar to the lower averaged rates, which slightly changes as the averaged rate increases its value.

\subsubsection{Dependence on $\alpha$}
\begin{figure}[h]
    \includegraphics[width=0.9\linewidth]{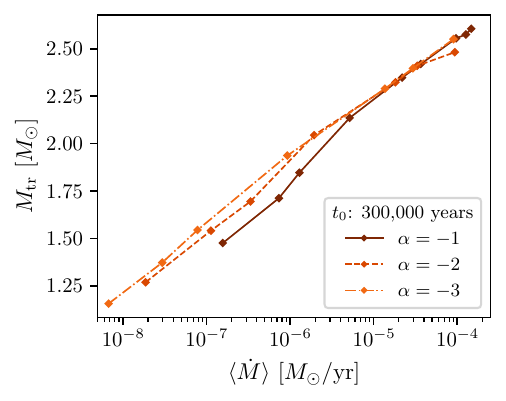}
    \caption{Results for the models with a power-law accretion rate, as defined in Eq. \eqref{Mpower}, assuming $t_0=300,000$ years and different power-law indexes, for a solar metallicity. Here, we show the transition mass for which at least $50\%$ of the protostar has become radiative as a function of the average accretion rate defined via Eq.\eqref{maverage}.}
    \label{powerlaw:different_alphas}
\end{figure}

In Figure \ref{powerlaw:different_alphas} the simulations for models with $\alpha=-1,-2,-3$ are shown assuming a fixed metallicity of $Z=0.02$. The values are also given in Table \ref{tab:summary_alpha_all}. 

For the same initial accretion rates, the average rate changes by more than an order of magnitude, considering values of $\alpha$ between $-1$ and $-3$. Hence, the greatest changes occur at the lower averaged accretion rates, while for the higher ones the curves behave similarly between different values of $\alpha$, reaching values for the transition mass of $2.61, 2.48 $ and $2.55 \ {M}_{\odot}$ for $\alpha=-1,-2,-3$, respectively.

\subsection{Comparison of all the models}

\begin{figure}[h]
   
    \includegraphics[width=0.48\textwidth]{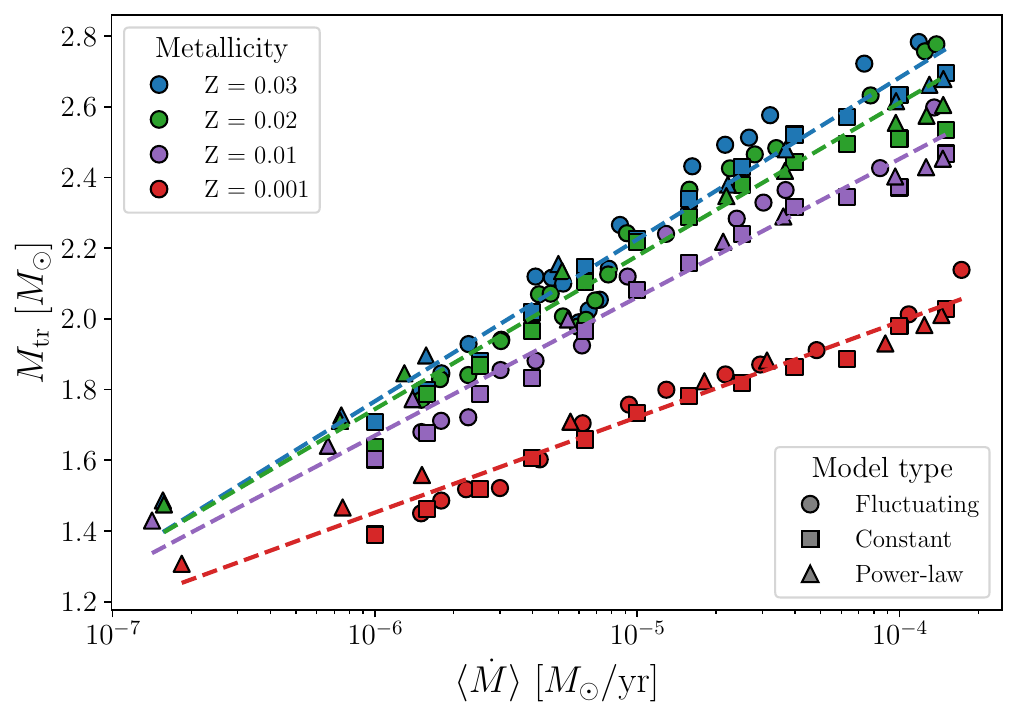}
    \caption{Overlay of the results from the constant, oscillating and power-law accretion scenarios, showing the behavior of the transition mass for three different prescriptions of the accretion rate, comparing them with four different metallicities. The lines show the fits to the four different groups with the parameters given in Table \ref{tab:combined_fits}.}
    \label{all_simulations}
\end{figure}

A comparison between the data shown in Figures \ref{constant}, \ref{fig:difz}, and \ref{powerlaw:constant_alpha} is presented in Figure \ref{all_simulations}, which includes the results obtained for the different accretion histories and metallicities. {As shown above}, metallicity plays a key role in determining the transition between radiative and convective states, whereas {for our adopted polytropic initial models and the accretion rate histories imposed after initialization}, the accretion history appears to have a comparatively limited influence.

{Comparing} accretion luminosity with nuclear luminosity at the end of {the calculation}, we {find} that the former {exceeds} the latter {at} high accretion rates. {Consequently}, some of the models do not reach the classical ZAMS at the end. By contrast, {models with} lower accretion rate have $ L_{nuc} > L_{acc}$ at the end {of the calculation}, {and reach} the classical ZAMS. For the purposes of this study, {however, it is not essential} whether the ZAMS is reached; our main goal is to {characterize} and quantify the transition {at which} the protostar becomes predominantly radiative. Nevertheless, the subsequent long-term evolution may depend on the evolution of the mass accretion rates {and, for stars whose accretion ceases before the ZAMS, on the pre-main-sequence contraction at fixed mass (see Sect.~4)}.

\subsection{Additional uncertainties}
A relevant additional uncertainty concerns the question of whether the full protostar becomes radiative before the end of the calculation. To assess this question, the Kippenhahn diagrams for the cases with a constant accretion rate are provided as supplementary material on Zenodo (see the Data availability section). We find that, particularly for models with metallicities of at least $Z=0.01$ and accretion rates of $4\cdot10^{-6}$~$ M_\odot$~yr$^{-1}$, many of the models still retain an outer convective envelope until they are close to the stopping point. The presence of such an envelope should be investigated further with 3D magneto-hydrodynamical models. We further checked the robustness of our criterion for the transition mass, considering different percentages at which the transition is evaluated (see Appendix~\ref{robust}). There is in general a small systematic shift when increasing the percentage, as the protostar needs to evolve further to fulfill the criterion. In addition, we emphasize that some models do not reach fractions above $\sim75\%$ of the mass as high accretion rates tend to favor convection within the envelope.

\section{Discussion and conclusions}\label{discusionandconclusion}

We have presented a series of stellar models using the MESA code to investigate the pre-main-sequence evolution of intermediate-mass stars under different assumptions for the time-evolution of the accretion rate and the metallicity. In particular, we have considered scenarios with constant, {oscillating}, and power-law accretion rates. For metallicity, we have explored a parameter space from $Z = 0.001$ to $0.03$. Our main goal was to determine the mass at which the transition occurs from a mostly convective to a mostly radiative state, which we quantified and plotted as a function of the average accretion rate during the protostellar evolution. We generally found that this transition occurs between a minimum mass of $\sim1.3$$M_\odot$ and a maximum mass of $\sim2.8$$M_\odot$, {in our reference models ($M_{\rm in}=1\,M_\odot$); for other initial masses, $M_{\rm tr}$ reaches up to $\sim4\,M_\odot$ at $\dot M\approx10^{-4}\,M_\odot\,\mathrm{yr}^{-1}$ (Appendix \ref{apendixg})}. The masses for the transition weakly depend on the average accretion rate and tend to increase with it. A clear dependence on the metallicity has also been found.

We explored to what extent the time evolution of the accretion rate is relevant. As a reference case, we first studied the constant accretion scenario. As shown in Table \ref{tab:summary_constant}, the time at which the protostar reaches the transition mass is always shorter for lower metallicities; this is also true for the transition mass itself, which decreases with metallicity. We also note that as the constant accretion rate increases, the values of the transition mass also increase for each metallicity, and are well described by a logarithmic fit (see Table \ref{tab:combined_fits}).

We further {considered} the effects of {oscillating} accretion rates, {with periods} between $10,000$ and $50,000$ years,  motivated by episodic accretion scenarios. In this case, we found a mostly weak dependence on the accretion period, which only occasionally influenced the point when the transition occurred in cases of somewhat longer periods. The main dependence also in this case was with respect to the average accretion rate and the metallicity. {Moreover}, our last case of a decreasing power-law accretion rate exhibits essentially the same behavior. This suggests that, to first order, the time evolution of the accretion {rate does not significantly affect the transitions,} {at least for the adopted polytropic initial models and the accretion histories imposed after initialization}, but the average accretion rate and metallicity predominantly govern the point where the transition to a radiative state {occurs}.

In our {reference} models, the transition mass slightly increases with the accretion rate under all assumptions, but only by about a solar mass, while the average accretion rate varied by 2-3 orders of magnitude. This means that the transition to the radiative state usually occurs at a somewhat similar point, {between $\sim1.3$ and $2.8\,M_\odot$}. After that transition, magnetic fields are expected to survive more easily{, provided that the radiative structure is maintained until the ZAMS. This could provide} a potential explanation for the larger fraction of stars with magnetic fields at high stellar masses. Our results suggest that also a dependence on metallicity should be present, implying lower transition masses for lower metallicity, which should translate into larger fractions of stars with strong magnetic fields. Potentially, this prediction can be tested by analyzing the observed samples.

An important caveat concerns the fact that protostars with higher accretion rates and metallicities {$Z\geq0.01$} tend to still have a convective outer envelope {at the end of the calculation}. It is possible that in these cases, the accreted magnetic field could be dissipated via convection, even if such a case so far has not been investigated via 3D magneto-hydrodynamical simulations. It is then possible that the observed radiative stars originate predominantly from cases of lower accretion rates, even if it is {not possible to draw} a definite conclusion. We also emphasize that within this study, we have adopted several {simplifications} for the accretion rate. Preferably, though, it would be desirable to calculate the protostellar evolution considering a sample of time-dependent accretion rate from 3D simulations of intermediate-mass star formation, while so far star formation studies focus predominantly on the low-mass regime \citep[e.g.][]{Riaz2018,Riaz2021,Wurster2022} or the high-mass regime \citep[e.g.][]{Peters2010,Seifried2013}.

We have neglected the effect of rotation in the models presented here, which can potentially affect the evolution of the internal structure of the protostar and somewhat change the mass scale for the transition. A general limitation is that the MESA code is a 1D model of stellar evolution. Since effects such as rotation and protostellar convection are intrinsically 3D phenomena, and MESA uses simplified recipes based on mixing-length theory, some discrepancies may arise. The presence of the fossil field itself may further alter the convection in the protostars, leading to a slightly different evolution. Similarly, the assumption of cold accretion is {an idealization}, and more violent accretion events could occasionally also lead to somewhat extended protostellar radii including a possible convective phase. In the long term, it will be important to verify stellar models through detailed 3D simulations. Since many aspects of protostellar evolution {are sensitive to} the adopted opacities,{ exploring the effects} of different opacity tables will be important for assessing the robustness of the predicted internal structure and the transition from the convective to the radiative phase in future studies.
{Since our analysis is restricted} to a radiative state across different models, it is important to {emphasize} that further exploration of the survival and general interaction between magnetic fields and the internal structure of the star is still required. Although {previous} studies {have addressed} the survival of magnetic fields in radiative stars, such as \citet{Braithwaite2004}, systematic investigations are {still needed} to fully understand {the conditions under which these fields can persist and shed light on their origin}. 

{A further limitation concerns the evolution after accretion ceases. Ap/Bp stars have final masses of about $1.5$--$6\,M_\odot$, so accretion must stop before the ZAMS, followed by pre-main-sequence contraction at fixed mass. We do not consider this phase here, since our models are stopped at the adopted stopping point. The internal structure may change and deep convection may reappear during a subsequent evolutionary stage. If that were to happen, it could affect the possible survival of fossil fields. Our conclusions on the survival of fossil fields therefore refer to the accretion phase, and a dedicated study of the post-accretion evolution is needed to confirm them.}

\section*{Data availability}
MESA files, including inlists, data files, the code used to reproduce the models and figures and the Kippenhahn diagrams for the constant accretion rate scenario are available on Zenodo: \href{https://doi.org/10.5281/zenodo.17652487}{doi:10.5281/zenodo.17652487}. 

\begin{acknowledgements}
DRGS and LV gratefully acknowledge the support of the ANID BASAL project FB21003. DRGS thanks the Alexander von Humboldt Foundation, Bonn, Germany, for the funding. J.A.R. acknowledges grant Vicerrectoría de Investigación y Desarrollo (VRID) Postdoctoral DICA N181/23.
\end{acknowledgements}

\bibliographystyle{aa} 
\bibliography{citas}  

@ARTICLE{Wurster2022,
       author = {{Wurster}, James and {Bate}, Matthew R. and {Price}, Daniel J. and {Bonnell}, Ian A.},
        title = "{On the origin of magnetic fields in stars - II. The effect of numerical resolution}",
      journal = {\mnras},
         year = 2022,
        month = mar,
       volume = {511},
       number = {1},
        pages = {746-764},
          doi = {10.1093/mnras/stac123},
archivePrefix = {arXiv},
       eprint = {2201.07253},
 primaryClass = {astro-ph.SR},
       adsurl = {https://ui.adsabs.harvard.edu/abs/2022MNRAS.511..746W}
}

@ARTICLE{Seifried2013,
       author = {{Seifried}, D. and {Banerjee}, R. and {Pudritz}, R.~E. and {Klessen}, R.~S.},
        title = "{Turbulence-induced disc formation in strongly magnetized cloud cores}",
      journal = {\mnras},
         year = 2013,
        month = jul,
       volume = {432},
       number = {4},
        pages = {3320-3331},
          doi = {10.1093/mnras/stt682},
archivePrefix = {arXiv},
       eprint = {1302.4901},
 primaryClass = {astro-ph.GA},
       adsurl = {https://ui.adsabs.harvard.edu/abs/2013MNRAS.432.3320S}
}

@ARTICLE{Peters2010,
       author = {{Peters}, Thomas and {Klessen}, Ralf S. and {Mac Low}, Mordecai-Mark and {Banerjee}, Robi},
        title = "{Limiting Accretion onto Massive Stars by Fragmentation-induced Starvation}",
      journal = {\apj},
         year = 2010,
        month = dec,
       volume = {725},
       number = {1},
        pages = {134-145},
          doi = {10.1088/0004-637X/725/1/134},
archivePrefix = {arXiv},
       eprint = {1005.3271},
 primaryClass = {astro-ph.GA},
       adsurl = {https://ui.adsabs.harvard.edu/abs/2010ApJ...725..134P}
}

@ARTICLE{Moss1989,
       author = {{Moss}, David},
        title = "{The origin and internal structure of the magnetic fields of the CP stars.}",
      journal = {\mnras},
         year = 1989,
        month = feb,
       volume = {236},
        pages = {629-644},
          doi = {10.1093/mnras/236.3.629},
       adsurl = {https://ui.adsabs.harvard.edu/abs/1989MNRAS.236..629M}
}

@ARTICLE{Riaz2018,
       author = {{Riaz}, R. and {Vanaverbeke}, S. and {Schleicher}, D.~R.~G.},
        title = "{Episodic accretion in binary protostars emerging from self-gravitating solar mass cores}",
      journal = {\aap},
         year = 2018,
        month = jun,
       volume = {614},
          eid = {A53},
        pages = {A53},
          doi = {10.1051/0004-6361/201732076},
archivePrefix = {arXiv},
       eprint = {1712.09646},
 primaryClass = {astro-ph.SR},
       adsurl = {https://ui.adsabs.harvard.edu/abs/2018A&A...614A..53R}
}

@ARTICLE{Riaz2021,
       author = {{Riaz}, R. and {Schleicher}, D.~R.~G. and {Vanaverbeke}, S. and {Klessen}, Ralf S.},
        title = "{Turbulence and its connection to episodic accretion in binary YSOs}",
      journal = {\mnras},
         year = 2021,
        month = nov,
       volume = {507},
       number = {4},
        pages = {6061-6077},
          doi = {10.1093/mnras/stab2489},
archivePrefix = {arXiv},
       eprint = {2108.12328},
 primaryClass = {astro-ph.SR},
       adsurl = {https://ui.adsabs.harvard.edu/abs/2021MNRAS.507.6061R}
}

@ARTICLE{Brandenburg2005,
       author = {{Brandenburg}, Axel and {Subramanian}, Kandaswamy},
        title = "{Astrophysical magnetic fields and nonlinear dynamo theory}",
      journal = {\physrep},
         year = 2005,
        month = oct,
       volume = {417},
       number = {1-4},
        pages = {1-209},
          doi = {10.1016/j.physrep.2005.06.005},
archivePrefix = {arXiv},
       eprint = {astro-ph/0405052},
 primaryClass = {astro-ph},
       adsurl = {https://ui.adsabs.harvard.edu/abs/2005PhR...417....1B}
}

@ARTICLE{Kapyla2023,
       author = {{K{\"a}pyl{\"a}}, Petri J. and {Browning}, Matthew K. and {Brun}, Allan Sacha and {Guerrero}, Gustavo and {Warnecke}, J{\"o}rn},
        title = "{Simulations of Solar and Stellar Dynamos and Their Theoretical Interpretation}",
      journal = {\ssr},
         year = 2023,
        month = oct,
       volume = {219},
       number = {7},
          eid = {58},
        pages = {58},
          doi = {10.1007/s11214-023-01005-6},
archivePrefix = {arXiv},
       eprint = {2305.16790},
 primaryClass = {astro-ph.SR},
       adsurl = {https://ui.adsabs.harvard.edu/abs/2023SSRv..219...58K}
}

@ARTICLE{Paxton2011,
       author = {{Paxton}, Bill and {Bildsten}, Lars and {Dotter}, Aaron and {Herwig}, Falk and {Lesaffre}, Pierre and {Timmes}, Frank},
        title = "{Modules for Experiments in Stellar Astrophysics (MESA)}",
      journal = {\apjs},
         year = 2011,
        month = jan,
       volume = {192},
       number = {1},
          eid = {3},
        pages = {3},
          doi = {10.1088/0067-0049/192/1/3},
archivePrefix = {arXiv},
       eprint = {1009.1622},
 primaryClass = {astro-ph.SR},
       adsurl = {https://ui.adsabs.harvard.edu/abs/2011ApJS..192....3P}
}

@ARTICLE{Paxton2013,
       author = {{Paxton}, Bill and {Cantiello}, Matteo and {Arras}, Phil and {Bildsten}, Lars and {Brown}, Edward F. and {Dotter}, Aaron and {Mankovich}, Christopher and {Montgomery}, M.~H. and {Stello}, Dennis and {Timmes}, F.~X. and {Townsend}, Richard},
        title = "{Modules for Experiments in Stellar Astrophysics (MESA): Planets, Oscillations, Rotation, and Massive Stars}",
      journal = {\apjs},
         year = 2013,
        month = sep,
       volume = {208},
       number = {1},
          eid = {4},
        pages = {4},
          doi = {10.1088/0067-0049/208/1/4},
archivePrefix = {arXiv},
       eprint = {1301.0319},
 primaryClass = {astro-ph.SR},
       adsurl = {https://ui.adsabs.harvard.edu/abs/2013ApJS..208....4P}
}

@ARTICLE{Paxton2015,
       author = {{Paxton}, Bill and {Marchant}, Pablo and {Schwab}, Josiah and {Bauer}, Evan B. and {Bildsten}, Lars and {Cantiello}, Matteo and {Dessart}, Luc and {Farmer}, R. and {Hu}, H. and {Langer}, N. and {Townsend}, R.~H.~D. and {Townsley}, Dean M. and {Timmes}, F.~X.},
        title = "{Modules for Experiments in Stellar Astrophysics (MESA): Binaries, Pulsations, and Explosions}",
      journal = {\apjs},
         year = 2015,
        month = sep,
       volume = {220},
       number = {1},
          eid = {15},
        pages = {15},
          doi = {10.1088/0067-0049/220/1/15},
archivePrefix = {arXiv},
       eprint = {1506.03146},
 primaryClass = {astro-ph.SR},
       adsurl = {https://ui.adsabs.harvard.edu/abs/2015ApJS..220...15P}
}

@ARTICLE{Paxton2019,
       author = {{Paxton}, Bill and {Smolec}, R. and {Schwab}, Josiah and {Gautschy}, A. and {Bildsten}, Lars and {Cantiello}, Matteo and {Dotter}, Aaron and {Farmer}, R. and {Goldberg}, Jared A. and {Jermyn}, Adam S. and {Kanbur}, S.~M. and {Marchant}, Pablo and {Thoul}, Anne and {Townsend}, Richard H.~D. and {Wolf}, William M. and {Zhang}, Michael and {Timmes}, F.~X.},
        title = "{Modules for Experiments in Stellar Astrophysics (MESA): Pulsating Variable Stars, Rotation, Convective Boundaries, and Energy Conservation}",
      journal = {\apjs},
         year = 2019,
        month = jul,
       volume = {243},
       number = {1},
          eid = {10},
        pages = {10},
          doi = {10.3847/1538-4365/ab2241},
archivePrefix = {arXiv},
       eprint = {1903.01426},
 primaryClass = {astro-ph.SR},
       adsurl = {https://ui.adsabs.harvard.edu/abs/2019ApJS..243...10P}
}

@ARTICLE{Choi2016,
       author = {{Choi}, Jieun and {Dotter}, Aaron and {Conroy}, Charlie and {Cantiello}, Matteo and {Paxton}, Bill and {Johnson}, Benjamin D.},
        title = "{Mesa Isochrones and Stellar Tracks (MIST). I. Solar-scaled Models}",
      journal = {\apj},
         year = 2016,
        month = jun,
       volume = {823},
       number = {2},
          eid = {102},
        pages = {102},
          doi = {10.3847/0004-637X/823/2/102},
archivePrefix = {arXiv},
       eprint = {1604.08592},
 primaryClass = {astro-ph.SR},
       adsurl = {https://ui.adsabs.harvard.edu/abs/2016ApJ...823..102C}
}

@ARTICLE{Henyey1965,
       author = {{Henyey}, Louis and {Vardya}, M.~S. and {Bodenheimer}, Peter},
        title = "{Studies in Stellar Evolution. III. The Calculation of Model Envelopes.}",
      journal = {\apj},
         year = 1965,
        month = oct,
       volume = {142},
        pages = {841},
          doi = {10.1086/148357},
       adsurl = {https://ui.adsabs.harvard.edu/abs/1965ApJ...142..841H}
}

@ARTICLE{Rogers2002,
       author = {{Rogers}, F.~J. and {Nayfonov}, A.},
        title = "{Updated and Expanded OPAL Equation-of-State Tables: Implications for Helioseismology}",
      journal = {\apj},
         year = 2002,
        month = sep,
       volume = {576},
       number = {2},
        pages = {1064-1074},
          doi = {10.1086/341894},
       adsurl = {https://ui.adsabs.harvard.edu/abs/2002ApJ...576.1064R}
}

@ARTICLE{Saumon1995,
       author = {{Saumon}, D. and {Chabrier}, G. and {van Horn}, H.~M.},
        title = "{An Equation of State for Low-Mass Stars and Giant Planets}",
      journal = {\apjs},
         year = 1995,
        month = aug,
       volume = {99},
        pages = {713},
          doi = {10.1086/192204},
       adsurl = {https://ui.adsabs.harvard.edu/abs/1995ApJS...99..713S}
}

@ARTICLE{Blouin2020,
       author = {{Blouin}, Simon and {Shaffer}, Nathaniel R. and {Saumon}, Didier and {Starrett}, Charles E.},
        title = "{New Conductive Opacities for White Dwarf Envelopes}",
      journal = {\apj},
         year = 2020,
        month = aug,
       volume = {899},
       number = {1},
          eid = {46},
        pages = {46},
          doi = {10.3847/1538-4357/ab9e75},
archivePrefix = {arXiv},
       eprint = {2006.16390},
 primaryClass = {astro-ph.SR},
       adsurl = {https://ui.adsabs.harvard.edu/abs/2020ApJ...899...46B}
}

@ARTICLE{Grevesse1998,
       author = {{Grevesse}, N. and {Sauval}, A.~J.},
        title = "{Standard Solar Composition}",
      journal = {\ssr},
         year = 1998,
        month = may,
       volume = {85},
        pages = {161-174},
          doi = {10.1023/A:1005161325181},
       adsurl = {https://ui.adsabs.harvard.edu/abs/1998SSRv...85..161G}
}

@ARTICLE{Ferguson2005,
       author = {{Ferguson}, Jason W. and {Alexander}, David R. and {Allard}, France and {Barman}, Travis and {Bodnarik}, Julia G. and {Hauschildt}, Peter H. and {Heffner-Wong}, Amanda and {Tamanai}, Akemi},
        title = "{Low-Temperature Opacities}",
      journal = {\apj},
         year = 2005,
        month = apr,
       volume = {623},
       number = {1},
        pages = {585-596},
          doi = {10.1086/428642},
archivePrefix = {arXiv},
       eprint = {astro-ph/0502045},
 primaryClass = {astro-ph},
       adsurl = {https://ui.adsabs.harvard.edu/abs/2005ApJ...623..585F}
}

@ARTICLE{Palla1991,
       author = {{Palla}, Francesco and {Stahler}, Steven W.},
        title = "{The Evolution of Intermediate-Mass Protostars. I. Basic Results}",
      journal = {\apj},
         year = 1991,
        month = jul,
       volume = {375},
        pages = {288},
          doi = {10.1086/170188},
       adsurl = {https://ui.adsabs.harvard.edu/abs/1991ApJ...375..288P}
}

@ARTICLE{Palla1992,
       author = {{Palla}, Francesco and {Stahler}, Steven W.},
        title = "{The Evolution of Intermediate-Mass Protostars. II. Influence of the Accretion Flow}",
      journal = {\apj},
         year = 1992,
        month = jun,
       volume = {392},
        pages = {667},
          doi = {10.1086/171468},
       adsurl = {https://ui.adsabs.harvard.edu/abs/1992ApJ...392..667P}
}

@ARTICLE{Hosokawa2010,
       author = {{Hosokawa}, Takashi and {Yorke}, Harold W. and {Omukai}, Kazuyuki},
        title = "{Evolution of Massive Protostars Via Disk Accretion}",
      journal = {\apj},
         year = 2010,
        month = sep,
       volume = {721},
       number = {1},
        pages = {478-492},
          doi = {10.1088/0004-637X/721/1/478},
archivePrefix = {arXiv},
       eprint = {1005.2827},
 primaryClass = {astro-ph.SR},
       adsurl = {https://ui.adsabs.harvard.edu/abs/2010ApJ...721..478H}
}

@ARTICLE{Haemmerle2013,
       author = {{Haemmerl{\'e}}, L. and {Eggenberger}, P. and {Meynet}, G. and {Maeder}, A. and {Charbonnel}, C.},
        title = "{Star formation with disc accretion and rotation. I. Stars between 2 and 22 M$_{{\ensuremath{\odot}}}$ at solar metallicity}",
      journal = {\aap},
         year = 2013,
        month = sep,
       volume = {557},
          eid = {A112},
        pages = {A112},
          doi = {10.1051/0004-6361/201321359},
archivePrefix = {arXiv},
       eprint = {1306.5871},
 primaryClass = {astro-ph.SR},
       adsurl = {https://ui.adsabs.harvard.edu/abs/2013A&A...557A.112H}
}

@ARTICLE{Jermyn2023,
       author = {{Jermyn}, Adam S. and {Bauer}, Evan B. and {Schwab}, Josiah and {Farmer}, R. and {Ball}, Warrick H. and {Bellinger}, Earl P. and {Dotter}, Aaron and {Joyce}, Meridith and {Marchant}, Pablo and {Mombarg}, Joey S.~G. and {Wolf}, William M. and {Sunny Wong}, Tin Long and {Cinquegrana}, Giulia C. and {Farrell}, Eoin and {Smolec}, R. and {Thoul}, Anne and {Cantiello}, Matteo and {Herwig}, Falk and {Toloza}, Odette and {Bildsten}, Lars and {Townsend}, Richard H.~D. and {Timmes}, F.~X.},
        title = "{Modules for Experiments in Stellar Astrophysics (MESA): Time-dependent Convection, Energy Conservation, Automatic Differentiation, and Infrastructure}",
      journal = {\apjs},
         year = 2023,
        month = mar,
       volume = {265},
       number = {1},
          eid = {15},
        pages = {15},
          doi = {10.3847/1538-4365/acae8d},
archivePrefix = {arXiv},
       eprint = {2208.03651},
 primaryClass = {astro-ph.SR},
       adsurl = {https://ui.adsabs.harvard.edu/abs/2023ApJS..265...15J}
}

@ARTICLE{Stahler1980a,
       author = {{Stahler}, S.~W. and {Shu}, F.~H. and {Taam}, R.~E.},
        title = "{The evolution of protostars. I - Global formulation and results}",
      journal = {\apj},
         year = 1980,
        month = oct,
       volume = {241},
        pages = {637-654},
          doi = {10.1086/158377},
       adsurl = {https://ui.adsabs.harvard.edu/abs/1980ApJ...241..637S}
}

@ARTICLE{Baraffe1998,
       author = {{Baraffe}, I. and {Chabrier}, G. and {Allard}, F. and {Hauschildt}, P.~H.},
        title = "{Evolutionary models for solar metallicity low-mass stars: mass-magnitude relationships and color-magnitude diagrams}",
      journal = {\aap},
         year = 1998,
        month = sep,
       volume = {337},
        pages = {403-412},
          doi = {10.48550/arXiv.astro-ph/9805009},
archivePrefix = {arXiv},
       eprint = {astro-ph/9805009},
 primaryClass = {astro-ph},
       adsurl = {https://ui.adsabs.harvard.edu/abs/1998A&A...337..403B}
}

@ARTICLE{Palla2000,
       author = {{Palla}, Francesco and {Stahler}, Steven W.},
        title = "{Accelerating Star Formation in Clusters and Associations}",
      journal = {\apj},
         year = 2000,
        month = sep,
       volume = {540},
       number = {1},
        pages = {255-270},
          doi = {10.1086/309312},
       adsurl = {https://ui.adsabs.harvard.edu/abs/2000ApJ...540..255P}
}

@ARTICLE{Larson1969,
       author = {{Larson}, Richard B.},
        title = "{Numerical calculations of the dynamics of collapsing proto-star}",
      journal = {\mnras},
         year = 1969,
        month = jan,
       volume = {145},
        pages = {271},
          doi = {10.1093/mnras/145.3.271},
       adsurl = {https://ui.adsabs.harvard.edu/abs/1969MNRAS.145..271L}
}

@ARTICLE{McNally1964,
       author = {{McNally}, D.},
        title = "{The Collapse of Interstellar Gas Clouds. I. The Effect of Cooling on Clouds Having Initially Polytropic Density Distributions.}",
      journal = {\apj},
         year = 1964,
        month = oct,
       volume = {140},
        pages = {1088},
          doi = {10.1086/148007},
       adsurl = {https://ui.adsabs.harvard.edu/abs/1964ApJ...140.1088M}
}

@ARTICLE{Bodenheimer1968,
       author = {{Bodenheimer}, Peter and {Sweigart}, Allen},
        title = "{Dynamic Collapse of the Isothermal Sphere}",
      journal = {\apj},
         year = 1968,
        month = may,
       volume = {152},
        pages = {515},
          doi = {10.1086/149568},
       adsurl = {https://ui.adsabs.harvard.edu/abs/1968ApJ...152..515B}
}

@ARTICLE{Palla1990,
       author = {{Palla}, Francesco and {Stahler}, Steven W.},
        title = "{The Birthline for Intermediate-Mass Stars}",
      journal = {\apjl},
         year = 1990,
        month = sep,
       volume = {360},
        pages = {L47},
          doi = {10.1086/185809},
       adsurl = {https://ui.adsabs.harvard.edu/abs/1990ApJ...360L..47P}
}

@ARTICLE{Bohm-Vitense1958,
       author = {{B{\"o}hm-Vitense}, E.},
        title = "{{\"U}ber die Wasserstoffkonvektionszone in Sternen verschiedener Effektivtemperaturen und Leuchtkr{\"a}fte. Mit 5 Textabbildungen}",
      journal = {\zap},
         year = 1958,
        month = jan,
       volume = {46},
        pages = {108},
       adsurl = {https://ui.adsabs.harvard.edu/abs/1958ZA.....46..108B}
}

@article{Burbidge1957,
  title = {Synthesis of the Elements in Stars},
  author = {Burbidge, E. Margaret and Burbidge, G. R. and Fowler, William A. and Hoyle, F.},
  journal = {Rev. Mod. Phys.},
  volume = {29},
  issue = {4},
  pages = {547--650},
  numpages = {0},
  year = {1957},
  month = {Oct},
  publisher = {American Physical Society},
  doi = {10.1103/RevModPhys.29.547},
  url = {https://link.aps.org/doi/10.1103/RevModPhys.29.547}
}

@BOOK{Schwarzschild1958,
       author = {{Schwarzschild}, Martin},
        title = "{Structure and evolution of the stars.}",
         year = 1958,
       adsurl = {https://ui.adsabs.harvard.edu/abs/1958ses..book.....S}
}

@ARTICLE{Henyey1964,
       author = {{Henyey}, L.~G. and {Forbes}, J.~E. and {Gould}, N.~L.},
        title = "{A New Method of Automatic Computation of Stellar Evolution.}",
      journal = {\apj},
         year = 1964,
        month = jan,
       volume = {139},
        pages = {306},
          doi = {10.1086/147754},
       adsurl = {https://ui.adsabs.harvard.edu/abs/1964ApJ...139..306H}
}

@ARTICLE{Paczynski1970A,
       author = {{Paczy{\'n}ski}, B.},
        title = "{Evolution of Single Stars. I. Stellar Evolution from Main Sequence to White Dwarf or Carbon Ignition}",
      journal = {\actaa},
         year = 1970,
        month = jan,
       volume = {20},
        pages = {47},
       adsurl = {https://ui.adsabs.harvard.edu/abs/1970AcA....20...47P}
}

@article{Ekström2012,
	author = {{Ekström, S.} and {Georgy, C.} and {Eggenberger, P.} and {Meynet, G.} and {Mowlavi, N.} and {Wyttenbach, A.} and {Granada, A.} and {Decressin, T.} and {Hirschi, R.} and {Frischknecht, U.} and {Charbonnel, C.} and {Maeder, A.}},
	title = {Grids of stellar models with rotation - I. Models from 0.8 to 120 M⊙ at solar metallicity (Z = 0.014)⋆},
	DOI= "10.1051/0004-6361/201117751",
	url= "https://doi.org/10.1051/0004-6361/201117751",
	journal = {\aap},
	year = 2012,
	volume = 537,
	pages = "A146",
	month = "",
}

@ARTICLE{Eggleton1971,
       author = {{Eggleton}, Peter P.},
        title = "{The evolution of low mass stars}",
      journal = {\mnras},
         year = 1971,
        month = jan,
       volume = {151},
        pages = {351},
          doi = {10.1093/mnras/151.3.351},
       adsurl = {https://ui.adsabs.harvard.edu/abs/1971MNRAS.151..351E}
}

@ARTICLE{Brott2011,
       author = {{Brott}, I. and {de Mink}, S.~E. and {Cantiello}, M. and {Langer}, N. and {de Koter}, A. and {Evans}, C.~J. and {Hunter}, I. and {Trundle}, C. and {Vink}, J.~S.},
        title = "{Rotating massive main-sequence stars. I. Grids of evolutionary models and isochrones}",
      journal = {\aap},
         year = 2011,
        month = jun,
       volume = {530},
          eid = {A115},
        pages = {A115},
          doi = {10.1051/0004-6361/201016113},
archivePrefix = {arXiv},
       eprint = {1102.0530},
 primaryClass = {astro-ph.SR},
       adsurl = {https://ui.adsabs.harvard.edu/abs/2011A&A...530A.115B}
}

@ARTICLE{Georgy2013,
       author = {{Georgy}, C. and {Ekstr{\"o}m}, S. and {Eggenberger}, P. and {Meynet}, G. and {Haemmerl{\'e}}, L. and {Maeder}, A. and {Granada}, A. and {Groh}, J.~H. and {Hirschi}, R. and {Mowlavi}, N. and {Yusof}, N. and {Charbonnel}, C. and {Decressin}, T. and {Barblan}, F.},
        title = "{Grids of stellar models with rotation. III. Models from 0.8 to 120 M$_{{\ensuremath{\odot}}}$ at a metallicity Z = 0.002}",
      journal = {\aap},
         year = 2013,
        month = oct,
       volume = {558},
          eid = {A103},
        pages = {A103},
          doi = {10.1051/0004-6361/201322178},
archivePrefix = {arXiv},
       eprint = {1308.2914},
 primaryClass = {astro-ph.SR},
       adsurl = {https://ui.adsabs.harvard.edu/abs/2013A&A...558A.103G}
}

@article{Pols1995,
    author = {Pols, Omno R. and Tout, Christopher A. and Eggleton, Peter P. and Han, Zhanwen},
    title = {Approximate input physics for stellar modelling},
    journal = {MNRAS},
    volume = {274},
    number = {3},
    pages = {964-974},
    year = {1995},
    month = {06},
    issn = {0035-8711},
    doi = {10.1093/mnras/274.3.964},
    url = {https://doi.org/10.1093/mnras/274.3.964},
    eprint = {https://academic.oup.com/mnras/article-pdf/274/3/964/2968688/mnras274-0964.pdf},
}

@ARTICLE{Girardi2002,
       author = {{Girardi}, L. and {Bertelli}, G. and {Bressan}, A. and {Chiosi}, C. and {Groenewegen}, M.~A.~T. and {Marigo}, P. and {Salasnich}, B. and {Weiss}, A.},
        title = "{Theoretical isochrones in several photometric systems. I. Johnson-Cousins-Glass, HST/WFPC2, HST/NICMOS, Washington, and ESO Imaging Survey filter sets}",
      journal = {\aap},
         year = 2002,
        month = aug,
       volume = {391},
        pages = {195-212},
          doi = {10.1051/0004-6361:20020612},
archivePrefix = {arXiv},
       eprint = {astro-ph/0205080},
 primaryClass = {astro-ph},
       adsurl = {https://ui.adsabs.harvard.edu/abs/2002A&A...391..195G}
}

@article{Bressan2012,
    author = {Bressan, Alessandro and Marigo, Paola and Girardi, LÃ©o and Salasnich, Bernardo and Dal Cero, Claudia and Rubele, Stefano and Nanni, Ambra},
    title = {PARSEC: stellar tracks and isochrones with the PAdova and TRieste Stellar Evolution Code},
    journal = {MNRAS},
    volume = {427},
    number = {1},
    pages = {127-145},
    year = {2012},
    month = {11},
    issn = {0035-8711},
    doi = {10.1111/j.1365-2966.2012.21948.x},
    url = {https://doi.org/10.1111/j.1365-2966.2012.21948.x},
    eprint = {https://academic.oup.com/mnras/article-pdf/427/1/127/18231042/427-1-127.pdf},
}

@ARTICLE{Auriere2007,
       author = {{Auri{\`e}re}, M. and {Wade}, G.~A. and {Silvester}, J. and {Ligni{\`e}res}, F. and {Bagnulo}, S. and {Bale}, K. and {Dintrans}, B. and {Donati}, J.~F. and {Folsom}, C.~P. and {Gruberbauer}, M. and {Hui Bon Hoa}, A. and {Jeffers}, S. and {Johnson}, N. and {Landstreet}, J.~D. and {L{\`e}bre}, A. and {Lueftinger}, T. and {Marsden}, S. and {Mouillet}, D. and {Naseri}, S. and {Paletou}, F. and {Petit}, P. and {Power}, J. and {Rincon}, F. and {Strasser}, S. and {Toqu{\'e}}, N.},
        title = "{Weak magnetic fields in Ap/Bp stars. Evidence for a dipole field lower limit and a tentative interpretation of the magnetic dichotomy}",
      journal = {\aap},
         year = 2007,
        month = dec,
       volume = {475},
       number = {3},
        pages = {1053-1065},
          doi = {10.1051/0004-6361:20078189},
archivePrefix = {arXiv},
       eprint = {0710.1554},
 primaryClass = {astro-ph},
       adsurl = {https://ui.adsabs.harvard.edu/abs/2007A&A...475.1053A}
}

@ARTICLE{Cowling1945,
       author = {{Cowling}, T.~G.},
        title = "{On the Sun's general magnetic field}",
      journal = {\mnras},
         year = 1945,
        month = jan,
       volume = {105},
        pages = {166},
          doi = {10.1093/mnras/105.3.166},
       adsurl = {https://ui.adsabs.harvard.edu/abs/1945MNRAS.105..166C}
}

@INPROCEEDINGS{Power2007,
       author = {{Power}, J. and {Wade}, G.~A. and {Hanes}, D.~A. and {Aurier}, M. and {Silvester}, J.},
        title = "{Properties of a volume-limited sample of magnetic Ap/Bp stars}",
    booktitle = {Physics of Magnetic Stars},
         year = 2007,
       editor = {{Romanyuk}, I.~I. and {Kudryavtsev}, D.~O. and {Neizvestnaya}, O.~M. and {Shapoval}, V.~M.},
        month = jan,
        pages = {89-97},
          doi = {10.48550/arXiv.astro-ph/0612557},
archivePrefix = {arXiv},
       eprint = {astro-ph/0612557},
 primaryClass = {astro-ph},
       adsurl = {https://ui.adsabs.harvard.edu/abs/2007pms..conf...89P}
}

@ARTICLE{Mathys2008,
       author = {{Mathys}, G.},
        title = "{Extremely slowly rotating magnetic Ap stars: recent results}",
      journal = {Contributions of the Astronomical Observatory Skalnate Pleso},
         year = 2008,
        month = apr,
       volume = {38},
       number = {2},
        pages = {217-222},
       adsurl = {https://ui.adsabs.harvard.edu/abs/2008CoSka..38..217M}
}

@ARTICLE{Braithwaite2017,
       author = {{Braithwaite}, Jonathan and {Spruit}, Henk C.},
        title = "{Magnetic fields in non-convective regions of stars}",
      journal = {Royal Society Open Science},
         year = 2017,
        month = feb,
       volume = {4},
       number = {2},
          eid = {160271},
        pages = {160271},
          doi = {10.1098/rsos.160271},
archivePrefix = {arXiv},
       eprint = {1510.03198},
 primaryClass = {astro-ph.SR},
       adsurl = {https://ui.adsabs.harvard.edu/abs/2017RSOS....460271B}
}

@ARTICLE{Donati2009,
       author = {{Donati}, J. -F. and {Landstreet}, J.~D.},
        title = "{Magnetic Fields of Nondegenerate Stars}",
      journal = {\araa},
         year = 2009,
        month = sep,
       volume = {47},
       number = {1},
        pages = {333-370},
          doi = {10.1146/annurev-astro-082708-101833},
archivePrefix = {arXiv},
       eprint = {0904.1938},
 primaryClass = {astro-ph.SR},
       adsurl = {https://ui.adsabs.harvard.edu/abs/2009ARA&A..47..333D}
}

@ARTICLE{Alecian2013,
       author = {{Alecian}, E. and {Neiner}, C. and {Mathis}, S. and {Catala}, C. and {Kochukhov}, O. and {Landstreet}, J.},
        title = "{The dramatic change of the fossil magnetic field of HD 190073: evidence of the birth of the convective core in a Herbig star?}",
      journal = {\aap},
         year = 2013,
        month = jan,
       volume = {549},
          eid = {L8},
        pages = {L8},
          doi = {10.1051/0004-6361/201220796},
archivePrefix = {arXiv},
       eprint = {1301.1804},
 primaryClass = {astro-ph.SR},
       adsurl = {https://ui.adsabs.harvard.edu/abs/2013A&A...549L...8A}
}

@ARTICLE{Schleicher2023,
       author = {{Schleicher}, Dominik R.~G. and {Hidalgo}, Juan Pablo and {Galli}, Daniele},
        title = "{Survival of fossil fields during the pre-main sequence evolution of intermediate-mass stars}",
      journal = {\aap},
         year = 2023,
        month = oct,
       volume = {678},
          eid = {A204},
        pages = {A204},
          doi = {10.1051/0004-6361/202346809},
archivePrefix = {arXiv},
       eprint = {2306.17131},
 primaryClass = {astro-ph.SR},
       adsurl = {https://ui.adsabs.harvard.edu/abs/2023A&A...678A.204S}
}

@ARTICLE{Dudorov1990,
       author = {{Dudorov}, A.~E. and {Tutukov}, A.~V.},
        title = "{The Magnetic Field of CP Stars}",
      journal = {\sovast},
         year = 1990,
        month = apr,
       volume = {34},
        pages = {171},
       adsurl = {https://ui.adsabs.harvard.edu/abs/1990SvA....34..171D}
}

@ARTICLE{Dudorov2015,
       author = {{Dudorov}, Alexander E. and {Khaibrakhmanov}, Sergey A.},
        title = "{Theory of fossil magnetic field}",
      journal = {ASR},
         year = 2015,
        month = feb,
       volume = {55},
       number = {3},
        pages = {843-850},
          doi = {10.1016/j.asr.2014.05.034},
archivePrefix = {arXiv},
       eprint = {1406.4663},
 primaryClass = {astro-ph.SR},
       adsurl = {https://ui.adsabs.harvard.edu/abs/2015AdSpR..55..843D}
}

@ARTICLE{Braithwaite2004,
       author = {{Braithwaite}, Jonathan and {Spruit}, Hendrik C.},
        title = "{A fossil origin for the magnetic field in A stars and white dwarfs}",
      journal = {\nat},
         year = 2004,
        month = oct,
       volume = {431},
       number = {7010},
        pages = {819-821},
          doi = {10.1038/nature02934},
archivePrefix = {arXiv},
       eprint = {astro-ph/0502043},
 primaryClass = {astro-ph},
       adsurl = {https://ui.adsabs.harvard.edu/abs/2004Natur.431..819B}
}

@ARTICLE{Braithwaite2006,
       author = {{Braithwaite}, J. and {Nordlund}, {\r{A}}.},
        title = "{Stable magnetic fields in stellar interiors}",
      journal = {\aap},
         year = 2006,
        month = may,
       volume = {450},
       number = {3},
        pages = {1077-1095},
          doi = {10.1051/0004-6361:20041980},
archivePrefix = {arXiv},
       eprint = {astro-ph/0510316},
 primaryClass = {astro-ph},
       adsurl = {https://ui.adsabs.harvard.edu/abs/2006A&A...450.1077B}
}

@ARTICLE{Braithwaite2008,
       author = {{Braithwaite}, Jonathan},
        title = "{On non-axisymmetric magnetic equilibria in stars}",
      journal = {\mnras},
         year = 2008,
        month = jun,
       volume = {386},
       number = {4},
        pages = {1947-1958},
          doi = {10.1111/j.1365-2966.2008.13218.x},
archivePrefix = {arXiv},
       eprint = {0803.1661},
 primaryClass = {astro-ph},
       adsurl = {https://ui.adsabs.harvard.edu/abs/2008MNRAS.386.1947B}
}

@article{Becerra2022,
    author = {Becerra, Laura and Reisenegger, Andreas and Valdivia, Juan Alejandro and Gusakov, Mikhail},
    title = {Stability of axially symmetric magnetic fields in stars},
    journal = {MNRAS},
    volume = {517},
    number = {1},
    pages = {560-568},
    year = {2022},
    month = {09},
    issn = {0035-8711},
    doi = {10.1093/mnras/stac2704},
    url = {https://doi.org/10.1093/mnras/stac2704},
    eprint = {https://academic.oup.com/mnras/article-pdf/517/1/560/46356006/stac2704.pdf},
}

@ARTICLE{Browning2004,
       author = {{Browning}, Matthew K. and {Brun}, Allan Sacha and {Toomre}, Juri},
        title = "{Simulations of Core Convection in Rotating A-Type Stars: Differential Rotation and Overshooting}",
      journal = {\apj},
         year = 2004,
        month = jan,
       volume = {601},
       number = {1},
        pages = {512-529},
          doi = {10.1086/380198},
archivePrefix = {arXiv},
       eprint = {astro-ph/0310003},
 primaryClass = {astro-ph},
       adsurl = {https://ui.adsabs.harvard.edu/abs/2004ApJ...601..512B}
}

@ARTICLE{Preston1974,
       author = {{Preston}, G.~W.},
        title = "{The chemically peculiar stars of the upper main sequence.}",
      journal = {\araa},
         year = 1974,
        month = jan,
       volume = {12},
        pages = {257-277},
          doi = {10.1146/annurev.aa.12.090174.001353},
       adsurl = {https://ui.adsabs.harvard.edu/abs/1974ARA&A..12..257P}
}

@ARTICLE{Lecoanet2022,
       author = {{Lecoanet}, Daniel and {Bowman}, Dominic M. and {Van Reeth}, Timothy},
        title = "{Asteroseismic inference of the near-core magnetic field strength in the main-sequence B star HD 43317}",
      journal = {\mnras},
         year = 2022,
        month = may,
       volume = {512},
       number = {1},
        pages = {L16-L20},
          doi = {10.1093/mnrasl/slac013},
archivePrefix = {arXiv},
       eprint = {2202.03440},
 primaryClass = {astro-ph.SR},
       adsurl = {https://ui.adsabs.harvard.edu/abs/2022MNRAS.512L..16L}
}

@article{MacDonald2004,
    author = {MacDonald, James and Mullan, D. J.},
    title = {Magnetic fields in massive stars: dynamics and origin},
    journal = {MNRAS},
    volume = {348},
    number = {2},
    pages = {702-716},
    year = {2004},
    month = {02},
    issn = {0035-8711},
    doi = {10.1111/j.1365-2966.2004.07394.x},
    url = {https://doi.org/10.1111/j.1365-2966.2004.07394.x},
    eprint = {https://academic.oup.com/mnras/article-pdf/348/2/702/3079012/348-2-702.pdf},
}

@ARTICLE{Hidalgo2024,
       author = {{Hidalgo}, J.~P. and {K{\"a}pyl{\"a}}, P.~J. and {Schleicher}, D.~R.~G. and {Ortiz-Rodr{\'\i}guez}, C.~A. and {Navarrete}, F.~H.},
        title = "{Magnetohydrodynamic simulations of A-type stars: Long-term evolution of core dynamo cycles}",
      journal = {\aap},
         year = 2024,
        month = nov,
       volume = {691},
          eid = {A326},
        pages = {A326},
          doi = {10.1051/0004-6361/202449977},
archivePrefix = {arXiv},
       eprint = {2409.18066},
 primaryClass = {astro-ph.SR},
       adsurl = {https://ui.adsabs.harvard.edu/abs/2024A&A...691A.326H}
}

@article{MacGregor2003,
doi = {10.1086/346257},
url = {https://dx.doi.org/10.1086/346257},
year = {2003},
month = {mar},
publisher = {},
volume = {586},
number = {1},
pages = {480},
author = {MacGregor, K. B. and Cassinelli, J. P.},
title = {Magnetic Fields in Massive Stars. II. The Buoyant Rise of Magnetic Flux Tubes through the Radiative Interior},
journal = {ApJ}
}

@ARTICLE{Featherstone2009,
       author = {{Featherstone}, Nicholas A. and {Browning}, Matthew K. and {Brun}, Allan Sacha and {Toomre}, Juri},
        title = "{Effects of Fossil Magnetic Fields on Convective Core Dynamos in A-type Stars}",
      journal = {\apj},
         year = 2009,
        month = nov,
       volume = {705},
       number = {1},
        pages = {1000-1018},
          doi = {10.1088/0004-637X/705/1/1000},
       adsurl = {https://ui.adsabs.harvard.edu/abs/2009ApJ...705.1000F}
}

@ARTICLE{Vallee1997,
       author = {{Vallee}, Jacques P.},
        title = "{Observations of the Magnetic Fields Inside and Outside the Milky Way, Starting with Globules (\raisebox{-0.5ex}\textasciitilde 1 parsec), Filaments, Clouds, Superbubbles, Spiral Arms, Galaxies, Superclusters, and Ending with the Cosmological Universe's Background Surface (at \raisebox{-0.5ex}\textasciitilde 8 Teraparsecs)}",
      journal = {\fcp},
         year = 1997,
        month = jan,
       volume = {19},
        pages = {1-89},
       adsurl = {https://ui.adsabs.harvard.edu/abs/1997FCPh...19....1V}
}

@ARTICLE{Girart2006,
       author = {{Girart}, Josep M. and {Rao}, Ramprasad and {Marrone}, Daniel P.},
        title = "{Magnetic Fields in the Formation of Sun-Like Stars}",
      journal = {Science},
         year = 2006,
        month = aug,
       volume = {313},
       number = {5788},
        pages = {812-814},
          doi = {10.1126/science.1129093},
archivePrefix = {arXiv},
       eprint = {astro-ph/0609177},
 primaryClass = {astro-ph},
       adsurl = {https://ui.adsabs.harvard.edu/abs/2006Sci...313..812G}
}

@ARTICLE{Li2009,
       author = {{Li}, Hua-bai and {Dowell}, C. Darren and {Goodman}, Alyssa and {Hildebrand}, Roger and {Novak}, Giles},
        title = "{Anchoring Magnetic Field in Turbulent Molecular Clouds}",
      journal = {\apj},
         year = 2009,
        month = oct,
       volume = {704},
       number = {2},
        pages = {891-897},
          doi = {10.1088/0004-637X/704/2/891},
archivePrefix = {arXiv},
       eprint = {0908.1549},
 primaryClass = {astro-ph.GA},
       adsurl = {https://ui.adsabs.harvard.edu/abs/2009ApJ...704..891L}
}

@ARTICLE{Chapman2013,
       author = {{Chapman}, Nicholas L. and {Davidson}, Jacqueline A. and {Goldsmith}, Paul F. and {Houde}, Martin and {Kwon}, Woojin and {Li}, Zhi-Yun and {Looney}, Leslie W. and {Matthews}, Brenda and {Matthews}, Tristan G. and {Novak}, Giles and {Peng}, Ruisheng and {Vaillancourt}, John E. and {Volgenau}, Nikolaus H.},
        title = "{Alignment between Flattened Protostellar Infall Envelopes and Ambient Magnetic Fields}",
      journal = {\apj},
         year = 2013,
        month = jun,
       volume = {770},
       number = {2},
          eid = {151},
        pages = {151},
          doi = {10.1088/0004-637X/770/2/151},
archivePrefix = {arXiv},
       eprint = {1305.2922},
 primaryClass = {astro-ph.GA},
       adsurl = {https://ui.adsabs.harvard.edu/abs/2013ApJ...770..151C}
}

@ARTICLE{Crutcher2012,
       author = {{Crutcher}, Richard M.},
        title = "{Magnetic Fields in Molecular Clouds}",
      journal = {\araa},
         year = 2012,
        month = sep,
       volume = {50},
        pages = {29-63},
          doi = {10.1146/annurev-astro-081811-125514},
       adsurl = {https://ui.adsabs.harvard.edu/abs/2012ARA&A..50...29C}
}

@ARTICLE{Koley2022,
       author = {{Koley}, Atanu and {Roy}, Nirupam and {Momjian}, Emmanuel and {Sarma}, Anuj P. and {Datta}, Abhirup},
        title = "{Magnetic field measurement in TMC-1C using 22.3 GHz CCS Zeeman splitting}",
      journal = {\mnras},
         year = 2022,
        month = oct,
       volume = {516},
       number = {1},
        pages = {L48-L52},
          doi = {10.1093/mnrasl/slac085},
archivePrefix = {arXiv},
       eprint = {2207.12604},
 primaryClass = {astro-ph.GA},
       adsurl = {https://ui.adsabs.harvard.edu/abs/2022MNRAS.516L..48K}
}

@article{Desch2001,
doi = {10.1086/319703},
url = {https://dx.doi.org/10.1086/319703},
year = {2001},
month = {mar},
publisher = {},
volume = {550},
number = {1},
pages = {314},
author = {Desch, S. J. and Mouschovias, Telemachos Ch.},
title = {The Magnetic Decoupling Stage of Star Formation},
journal = {ApJ}
}

@ARTICLE{Nakano2002,
       author = {{Nakano}, Takenori and {Nishi}, Ryoichi and {Umebayashi}, Toyoharu},
        title = "{Mechanism of Magnetic Flux Loss in Molecular Clouds}",
      journal = {\apj},
         year = 2002,
        month = jul,
       volume = {573},
       number = {1},
        pages = {199-214},
          doi = {10.1086/340587},
archivePrefix = {arXiv},
       eprint = {astro-ph/0203223},
 primaryClass = {astro-ph},
       adsurl = {https://ui.adsabs.harvard.edu/abs/2002ApJ...573..199N}
}

@ARTICLE{Tomida2015,
       author = {{Tomida}, Kengo and {Okuzumi}, Satoshi and {Machida}, Masahiro N.},
        title = "{Radiation Magnetohydrodynamic Simulations of Protostellar Collapse: Nonideal Magnetohydrodynamic Effects and Early Formation of Circumstellar Disks}",
      journal = {\apj},
         year = 2015,
        month = mar,
       volume = {801},
       number = {2},
          eid = {117},
        pages = {117},
          doi = {10.1088/0004-637X/801/2/117},
archivePrefix = {arXiv},
       eprint = {1501.04102},
 primaryClass = {astro-ph.SR},
       adsurl = {https://ui.adsabs.harvard.edu/abs/2015ApJ...801..117T}
}

@ARTICLE{Masson2016,
       author = {{Masson}, J. and {Chabrier}, G. and {Hennebelle}, P. and {Vaytet}, N. and {Commer{\c{c}}on}, B.},
        title = "{Ambipolar diffusion in low-mass star formation. I. General comparison with the ideal magnetohydrodynamic case}",
      journal = {\aap},
         year = 2016,
        month = mar,
       volume = {587},
          eid = {A32},
        pages = {A32},
          doi = {10.1051/0004-6361/201526371},
archivePrefix = {arXiv},
       eprint = {1509.05630},
 primaryClass = {astro-ph.SR},
       adsurl = {https://ui.adsabs.harvard.edu/abs/2016A&A...587A..32M}
}

@article{Nakano_ubemayashi1988,
    author = {Nakano, Takenori and Umebayashi, Toyoharu},
    title = {Chapter 6. Effects of Magnetic Fields on Star Formation},
    journal = {Progress of Theoretical Physics Supplement},
    volume = {96},
    pages = {73-84},
    year = {1988},
    month = {03},
    issn = {0375-9687},
    doi = {10.1143/PTPS.96.73},
    url = {https://doi.org/10.1143/PTPS.96.73},
    eprint = {https://academic.oup.com/ptps/article-pdf/doi/10.1143/PTPS.96.73/5412831/96-73.pdf},
}

@INPROCEEDINGS{Nakano1988,
       author = {{Nakano}, T.},
        title = "{The Behaviour of Magnetic Fields in Star Forming Regions}",
    booktitle = {Galactic and Extragalactic Star Formation},
         year = 1988,
       editor = {{Pudritz}, Ralph E. and {Fich}, Michel},
       series = {NATO Advanced Study Institute (ASI) Series C},
       volume = {232},
        month = jan,
        pages = {111},
          doi = {10.1007/978-94-009-2973-9_6},
       adsurl = {https://ui.adsabs.harvard.edu/abs/1988ASIC..232..111N}
}

@article{Brun2005,
doi = {10.1086/430430},
url = {https://doi.org/10.1086/430430},
year = {2005},
month = {aug},
publisher = {},
volume = {629},
number = {1},
pages = {461},
author = {Brun, Allan Sacha and Browning, Matthew K. and Toomre, Juri},
title = {Simulations of Core Convection in Rotating A-Type Stars: Magnetic Dynamo Action},
journal = {ApJ}
}

@ARTICLE{Hidalgo2025,
       author = {{Hidalgo}, J.~P. and {K{\"a}pyl{\"a}}, P.~J. and {Schleicher}, D.~R.~G. and {Ortiz-Rodr{\'\i}guez}, C.~A. and {Navarrete}, F.~H.},
        title = "{Shaping core dynamos in A-type stars: The role of dipolar fossil fields}",
      journal = {\aap},
         year = 2025,
        month = jul,
       volume = {699},
          eid = {A250},
        pages = {A250},
          doi = {10.1051/0004-6361/202555151},
archivePrefix = {arXiv},
       eprint = {2506.01017},
 primaryClass = {astro-ph.SR},
       adsurl = {https://ui.adsabs.harvard.edu/abs/2025A&A...699A.250H}
}


\begin{appendix}
    
\section{Data for constant mass accretion rate scenario}

Table~\ref{tab:summary_constant}, summarizes the main results for the constant accretion rate scenario described in Section 3.1, considering four different metallicities.

\begin{table}[h]
\centering
\caption{Results from the constant mass accretion rate scenario.}
\label{tab:summary_constant}
\resizebox{\hsize}{!}{
\begin{tabular}{l lll lll}
\toprule
\ \ \ \ \ \ $\dot{ M}$ &
\multicolumn{3}{c}{$Z=0.001$} & 
\multicolumn{3}{c}{$Z=0.01$} \\
\cmidrule(lr){2-4} \cmidrule(lr){5-7}
[${M}_{\odot}\,\mathrm{yr}^{-1}$] &
$ M_{\mathrm{tr}}$ [${M}_{\odot}$] & $ t_{\mathrm{tr}}$ [Myr] & $ M_{\mathrm{stop}}$ [${M}_{\odot}$] &
$ M_{\mathrm{tr}}$ [${M}_{\odot}$] & $ t_{\mathrm{tr}}$ [Myr] & $ M_{\mathrm{stop}}$ [${M}_{\odot}$] \\
\midrule
$1.00\cdot10^{-6}$ & 1.39 & 0.39 & 2.93 & 1.60 & 0.60 & 3.21 \\
$1.58\cdot10^{-6}$ & 1.46 & 0.29 & 3.27 & 1.68 & 0.43 & 3.55 \\
$2.51\cdot10^{-6}$ & 1.52 & 0.21 & 3.67 & 1.79 & 0.31 & 3.92 \\
$3.98\cdot10^{-6}$ & 1.61 & 0.15 & 4.11 & 1.83 & 0.21 & 4.31 \\
$6.31\cdot10^{-6}$ & 1.66 & 0.10 & 4.60 & 1.97 & 0.15 & 4.76 \\
$1.00\cdot10^{-5}$ & 1.73 & 0.07 & 5.15 & 2.08 & 0.11 & 5.20 \\
$1.58\cdot10^{-5}$ & 1.78 & 0.05 & 5.75 & 2.16 & 0.07 & 3.62 \\
$2.51\cdot10^{-5}$ & 1.82 & 0.03 & 6.42 & 2.24 & 0.05 & 3.37\\
$3.98\cdot10^{-5}$ & 1.86 & 0.02 & 7.13 & 2.32 & 0.03 & 3.29 \\
$6.31\cdot10^{-5}$ & 1.89 & 0.01 & 7.88 & 2.35 & 0.02 & 3.24 \\
$1.00\cdot10^{-4}$ & 1.98 & 0.01 & 4.50 & 2.37 & 0.01 & 3.21 \\
$1.50\cdot10^{-4}$ & 2.03 & 0.01 & 4.31 & 2.47 & 0.01 & 3.19 \\
\bottomrule
\end{tabular}
}

\vspace{0.3cm}

\resizebox{\hsize}{!}{
\begin{tabular}{l lll lll}
\toprule
\ \ \ \ \ \ $\dot{ M}$ &
\multicolumn{3}{c}{$Z=0.02$} & 
\multicolumn{3}{c}{$Z=0.03$} \\
\cmidrule(lr){2-4} \cmidrule(lr){5-7}
[${M}_{\odot}\,\mathrm{yr}^{-1}$] &
$M_{\mathrm{tr}}$ [${M}_{\odot}$] & $t_{\mathrm{tr}}$ [Myr] & $ M_{\mathrm{stop}}$ [${M}_{\odot}$] &
$M_{\mathrm{tr}}$ [${M}_{\odot}$] & $t_{\mathrm{tr}}$ [Myr] & $ M_{\mathrm{stop}}$ [${M}_{\odot}$] \\
\midrule
$1.00\cdot10^{-6}$ & 1.64 & 0.64 & 3.43 & 1.71 & 0.71 & 2.96 \\
$1.58\cdot10^{-6}$ & 1.79 & 0.50 & 3.78 & 1.80 & 0.50 & 3.96 \\
$2.51\cdot10^{-6}$ & 1.87 & 0.35 & 4.19 & 1.88 & 0.35 & 4.39 \\
$3.98\cdot10^{-6}$ & 1.97 & 0.24 & 4.63 & 2.02 & 0.26 & 4.87 \\
$6.31\cdot10^{-6}$ & 2.11 & 0.18 & 5.11 & 2.15 & 0.18 & 3.09 \\
$1.00\cdot10^{-5}$ & 2.22 & 0.12 & 3.19 & 2.22 & 0.12 & 2.93 \\
$1.58\cdot10^{-5}$ & 2.29 & 0.08 & 3.06 & 2.34 & 0.08 & 2.86 \\
$2.51\cdot10^{-5}$ & 2.38 & 0.05 & 2.99 & 2.43 & 0.06 & 2.82 \\
$3.98\cdot10^{-5}$ & 2.44 & 0.04 & 2.96 & 2.52 & 0.04 & 2.79 \\
$6.31\cdot10^{-5}$ & 2.50 & 0.02 & 2.93 & 2.57 & 0.02 & 2.77 \\
$1.00\cdot10^{-4}$ & 2.51 & 0.02 & 2.91 & 2.63 & 0.02 & 2.76 \\
$1.50\cdot10^{-4}$ & 2.53 & 0.01 & 2.90 & 2.70 & 0.01 & 2.75 \\
\bottomrule
\end{tabular}
}
\tablefoot{
    These results correspond to the scenario described in Section \ref{constantrate}. Here, we show the constant accretion rate $\dot{  M}$, the mass and time of the transition $ M_{\mathrm{tr}}$ and $t_{\mathrm{\rm tr}}$, and the mass at the stopping condition $ M_{\mathrm{stop}}$ for four distinct metallicities.
}
\end{table}

\section{Fit to the data}
{To characterize} the relation between the transition mass and the mean mass accretion rate, {we fit the data with a logarithmic relation of the} form.
\begin{equation}\label{fiteqn}
    \ M_{\mathrm{tr}} = a \log_{10} \left( \langle \dot{M} \rangle \right) + b
\end{equation}
where $a$ and $b$ are free parameters {determined} from the fit. The values of the parameters obtained for the constant accretion scenario and the fit results for the entire dataset, including the different models,are summarized in Table \ref{tab:combined_fits}.

\begin{table}[h]
\centering
\caption{Fitting parameters using Eq.~\eqref{fiteqn} for different accretion scenarios.}
\label{tab:combined_fits}
\footnotesize 
\setlength{\tabcolsep}{8pt} 
\renewcommand{\arraystretch}{1.2}
\begin{tabular}{c cc cc}
\toprule
 & \multicolumn{2}{c}{Constant accretion} & \multicolumn{2}{c}{All prescriptions} \\
\cmidrule(lr){2-3} \cmidrule(lr){4-5} 
$Z$ & $a$ & $b$ & $a$ & $b$ \\
\midrule
0.03  & 0.4689 & 4.5462 & 0.4578 & 4.5138 \\
0.02  & 0.4226 & 4.2578 & 0.4331 & 4.3435 \\
0.01  & 0.4037 & 4.0497 & 0.3913 & 4.0178 \\
0.001 & 0.2822 & 3.1105 & 0.2701 & 3.0727 \\
\bottomrule
\end{tabular}
\tablefoot{
    The parameters $a$ and $b$ that best fit the data are shown as a function of the metallicity $Z$. The left section corresponds to the constant mass accretion rate scenario, while the right section corresponds to the three different prescriptions shown in Fig.~8.
}
\end{table}

\section{Data for the power-law mass accretion rate scenario}
In this appendix, we provide the main results for the power-law accretion rate scenario for different metallicities and considering variations of the power-law index.

\subsection{Simulations for metallicity variations}
In Table~\ref{tab:summary_powerlaw}, we summarize the results for the case of a power-law index $\alpha=-1$ varying the metallicity.

\begin{table}[h]
\centering
\caption{Results from the power-law mass accretion rate scenario.}
\label{tab:summary_powerlaw}
\resizebox{\hsize}{!}{
\begin{tabular}{l ccccc ccccc}
\toprule
\ \ \ \ \ \ $\dot{M}_0$ & 
\multicolumn{5}{c}{$Z=0.001$} & 
\multicolumn{5}{c}{$Z=0.01$} \\
\cmidrule(lr){2-6} \cmidrule(lr){7-11}
[${M}_{\odot}\,\mathrm{yr}^{-1}$] & 
$ M_{\mathrm{stop}}$ & $ t_{\mathrm{stop}}$& $\langle \dot{M} \rangle$ & $M_{\mathrm{tr}}$ & $t_{\mathrm{tr}}$ &
$ M_{\mathrm{stop}}$ & $ t_{\mathrm{stop}}$& $\langle \dot{M} \rangle$ & $M_{\mathrm{tr}}$ & $t_{\mathrm{tr}}$ \\
& 
[$\rm{M}_\odot$] & [Myr] & [${M}_{\odot}\,\mathrm{yr}^{-1}$] & [${M}_{\odot}$] & [Myr] & 
[${M}_{\odot}$] & [Myr] & [${M}_{\odot}\,\mathrm{yr}^{-1}$] & [${M}_{\odot}$] & [Myr] \\
\midrule
$1.00 \cdot 10^{-6}$ & 1.84 & 4.59 & $1.83 \cdot 10^{-7}$ & 1.31 & 0.53 & 1.95  & 6.74 & $1.41 \cdot 10^{-7}$ & 1.43 & 0.94 \\
$2.51 \cdot 10^{-6}$ & 2.56  & 2.07 & $7.53 \cdot 10^{-7}$ & 1.47 & 0.26 & 2.71  & 2.58 & $6.62 \cdot 10^{-7}$ & 1.64 & 0.40 \\
$3.98 \cdot 10^{-6}$ & 3.04  & 1.35 & $1.51 \cdot 10^{-6}$ & 1.56 & 0.18 & 3.19  & 1.58 & $1.39 \cdot 10^{-6}$ & 1.77 & 0.27 \\
$1.00 \cdot 10^{-5}$ & 4.24  & 0.58 & $5.57 \cdot 10^{-6}$ & 1.71 & 0.08 & 4.35  & 0.62 & $5.44 \cdot 10^{-6}$ & 2.00 & 0.12 \\
$2.51 \cdot 10^{-5}$ & 5.73  & 0.26 & $1.81 \cdot 10^{-5}$ & 1.82 & 0.03 & 3.43  & 0.11 & $2.13 \cdot 10^{-5}$ & 2.22 & 0.05 \\
$3.98 \cdot 10^{-5}$ & 6.56  & 0.18 & $3.13 \cdot 10^{-5}$ & 1.88 & 0.02 & 3.30  & 0.06 & $3.61 \cdot 10^{-5}$ & 2.29 & 0.03 \\
$1.00 \cdot 10^{-4}$ & 8.30  & 0.08 & $8.83 \cdot 10^{-5}$ & 1.93 & 0.01 & 3.21  & 0.02 & $9.64 \cdot 10^{-5}$ & 2.40 & 0.01 \\
$1.30 \cdot 10^{-4}$ & 4.38  & 0.03 & $1.25 \cdot 10^{-4}$ & 1.98 & 0.01 & 3.19  & 0.02 & $1.26 \cdot 10^{-4}$ & 2.43 & 0.01 \\
$1.50 \cdot 10^{-4}$ & 4.32  & 0.02 & $1.45 \cdot 10^{-4}$ & 2.01 & 0.01 & 3.19  & 0.02 & $1.46 \cdot 10^{-4}$ & 2.45 & 0.01 \\
\bottomrule
\end{tabular}
}

\vspace{0.3cm}

\resizebox{\hsize}{!}{
\begin{tabular}{l ccccc ccccc}
\toprule
\ \ \ \ \ \ $\dot{M}_0$  & 
\multicolumn{5}{c}{$Z=0.02$} & 
\multicolumn{5}{c}{$Z=0.03$} \\
\cmidrule(lr){2-6} \cmidrule(lr){7-11}
  [${M}_{\odot}\,\mathrm{yr}^{-1}$] & 
$ M_{\mathrm{stop}}$ & $ t_{\mathrm{stop}}$& $\langle \dot{M} \rangle$ & $M_{\mathrm{tr}}$ & $t_{\mathrm{tr}}$ &
$ M_{\mathrm{stop}}$ & $ t_{\mathrm{stop}}$& $\langle \dot{M} \rangle$ & $M_{\mathrm{tr}}$ & $t_{\mathrm{tr}}$ \\
& 
[${M}_{\odot}$] & [Myr] & [${M}_{\odot}\,\mathrm{yr}^{-1}$] & [${M}_{\odot}$] & [Myr] & 
[${M}_{\odot}$] & [Myr] & [${M}_{\odot}\,\mathrm{yr}^{-1}$] & [${M}_{\odot}$] & [Myr] \\
\midrule
$1.00 \cdot 10^{-6}$ & 1.91 & 5.80  & $1.57 \cdot 10^{-7}$ & 1.48 & 1.15 & 1.91 & 5.88  & $1.56 \cdot 10^{-7}$ & 1.49 & 1.20 \\
$2.51 \cdot 10^{-6}$ & 2.59 & 2.16  & $7.36 \cdot 10^{-7}$ & 1.71 & 0.47 & 2.58 & 2.12  & $7.45 \cdot 10^{-7}$ & 1.73 & 0.49 \\
$3.98 \cdot 10^{-6}$ & 3.33 & 1.80  & $1.30 \cdot 10^{-6}$ & 1.85 & 0.31 & 2.97 & 1.26  & $1.57 \cdot 10^{-6}$ & 1.90 & 0.33 \\
$1.00 \cdot 10^{-5}$ & 4.61 & 0.70 & $5.17 \cdot 10^{-6}$ & 2.14 & 0.14 & 4.78 & 0.75 & $5.01 \cdot 10^{-6}$ & 2.16 & 0.14 \\
$2.51 \cdot 10^{-5}$ & 3.02 & 0.09 & $2.19 \cdot 10^{-5}$ & 2.35 & 0.06 & 2.83 & 0.08 & $2.22 \cdot 10^{-5}$ & 2.38 & 0.06 \\
$3.98 \cdot 10^{-5}$ & 2.97 & 0.05 & $3.66 \cdot 10^{-5}$ & 2.42 & 0.04 & 2.79 & 0.05 & $3.69 \cdot 10^{-5}$ & 2.48 & 0.04 \\
$1.00 \cdot 10^{-4}$ & 2.92 & 0.02 & $9.68 \cdot 10^{-5}$ & 2.55 & 0.02 & 2.75 & 0.02 & $9.71 \cdot 10^{-5}$ & 2.62 & 0.02 \\
$1.30 \cdot 10^{-4}$ & 2.91 & 0.01 & $1.27 \cdot 10^{-4}$ & 2.57 & 0.01 & 2.75 & 0.01 & $1.30 \cdot 10^{-4}$ & 2.66 & 0.01 \\
$1.50 \cdot 10^{-4}$ & 2.91 & 0.01 & $1.47 \cdot 10^{-4}$ & 2.61 & 0.01 & 2.75 & 0.01 & $1.47 \cdot 10^{-4}$ & 2.68 & 0.01 \\
\bottomrule
\end{tabular}
}
\tablefoot{
     Scenario described in Section \ref{powerlaw}. We assumed a power-law index of $\alpha = -1$ and a $t_0 = 300,000$ years. Here are shown the initial accretion rate $\dot{M}_0$, the mass  $M_{\mathrm{stop}}$ and time $ t_{\mathrm{stop}}$ when the star reaches the stopping condition, the mean accretion rate of its evolution $\langle \dot{M} \rangle$ calculated according to Eq. \eqref{maverage}, the mass and time of the transition $M_{\mathrm{tr}}$ and $t_{\mathrm{tr}}$, respectively for four different metallicities.
}
\end{table}

\subsection{Simulations for power-law index variation}
In Table~\ref{tab:summary_alpha_all}, we summarize the results for the case of a solar metallicity varying the power-law index $\alpha$.

\begin{table}[h]
\centering
\caption{Results for the power-law mass accretion rate scenario.}
\label{tab:summary_alpha_all}
\resizebox{\hsize}{!}{
\begin{tabular}{l ccccc ccccc}
\toprule
\ \ \ \ \ \ $\dot{M}_0$ & 
\multicolumn{5}{c}{$\alpha = -3$} & 
\multicolumn{5}{c}{$\alpha = -2$} \\
\cmidrule(lr){2-6} \cmidrule(lr){7-11}
[${M}_{\odot}\,\mathrm{yr}^{-1}$] & 
$ M_{\mathrm{stop}}$ & $ t_{\mathrm{stop}}$& $\langle \dot{M} \rangle$ & $M_{\mathrm{tr}}$ & $t_{\mathrm{tr}}$ &
$ M_{\mathrm{stop}}$ & $ t_{\mathrm{stop}}$& $\langle \dot{M} \rangle$ & $M_{\mathrm{tr}}$ & $t_{\mathrm{tr}}$ \\
& 
[${M}_{\odot}$] & [Myr] & [${M}_{\odot}\,\mathrm{yr}^{-1}$] & [${M}_{\odot}$] & [Myr] & 
[${M}_{\odot}$] & [Myr] & [${M}_{\odot}\,\mathrm{yr}^{-1}$] & [${M}_{\odot}$] & [Myr] \\
\midrule
$1.00 \cdot 10^{-6}$ & 1.16 & 23.19 & $6.76 \cdot 10^{-9}$ & 1.16 & 3.35 & 1.30 & 16.16 & $1.87 \cdot 10^{-8}$ & 1.27 & 2.05 \\
$2.51 \cdot 10^{-6}$ & 1.39 & 13.07 & $2.96 \cdot 10^{-8}$ & 1.37 & 1.29 & 1.73 & 6.45  & $1.13 \cdot 10^{-7}$ & 1.54 & 0.73 \\
$3.98 \cdot 10^{-6}$ & 1.61 & 7.79  & $7.83 \cdot 10^{-8}$ & 1.54 & 0.64 & 2.11 & 3.29  & $3.36 \cdot 10^{-7}$ & 1.69 & 0.41 \\
$1.00 \cdot 10^{-5}$ & 2.48 & 1.59  & $9.27 \cdot 10^{-7}$ & 1.94 & 0.19 & 3.43 & 1.25  & $1.94 \cdot 10^{-6}$ & 2.04 & 0.16 \\
$2.51 \cdot 10^{-5}$ & 3.15 & 0.16  & $1.37 \cdot 10^{-5}$ & 2.29 & 0.07 & 3.06 & 0.11  & $1.82 \cdot 10^{-5}$ & 2.32 & 0.06 \\
$3.98 \cdot 10^{-5}$ & 2.99 & 0.07  & $2.95 \cdot 10^{-5}$ & 2.40 & 0.04 & 2.98 & 0.06  & $3.32 \cdot 10^{-5}$ & 2.41 & 0.04 \\
$1.00 \cdot 10^{-4}$ & 2.92 & 0.02  & $9.03 \cdot 10^{-5}$ & 2.55 & 0.02 & 2.92 & 0.02  & $9.36 \cdot 10^{-5}$ & 2.48 & 0.02 \\
\bottomrule
\end{tabular}
}
\tablefoot{
     Scenario described in Section \ref{powerlaw}. For these calculations, we assumed a metallicity of $Z=0.02$ and a $ t_0 = 300,000$ years. Here are shown the power-law indices $\alpha = -3$, $\alpha = -2$, with their corresponding stellar parameters. $M_{\rm stop}$ and $t_{\rm stop}$ refer to the stopping condition. Entries for $\alpha = -1$ are omitted because already shown in Table \ref{tab:summary_powerlaw}.
}
\end{table}

\section{Data for the oscillating mass accretion rate scenario}

\subsection{Simulations for variations of the period}
In Table~\ref{tab:summary_changing_periods}, we summarize the results for the case of a solar metallicity varying the period of the assumed oscillations.

\begin{table}[h]
\centering
\caption{Results for the oscillating mass accretion rate scenario (period variations).}
\label{tab:summary_changing_periods}
\resizebox{\hsize}{!}{
\begin{tabular}{c ccccc ccccc}
\toprule
Amplitude & \multicolumn{5}{c}{$P=10{,}000$ years} & \multicolumn{5}{c}{$P=25{,}000$ years} \\
\cmidrule(lr){2-6} \cmidrule(lr){7-11}
[${M}_{\odot}\,\mathrm{yr}^{-1}$] & $M_{\mathrm{stop}}$ & $t_{\mathrm{stop}}$ & $\langle \dot{M} \rangle $ & $M_{\mathrm{tr}}$ & $t_{\mathrm{tr}}$ &
$M_{\mathrm{stop}}$ & $t_{\mathrm{stop}}$ & $\langle \dot{M} \rangle$ & $M_{\mathrm{tr}}$ & $t_{\mathrm{tr}}$ \\
 & [${M}_{\odot}$] & [Myr] & [$10^{-6}{M}_{\odot}\,\mathrm{yr}^{-1}$] & [${M}_{\odot}$] & [Myr] & 
[${M}_{\odot}$] & [Myr] & [$10^{-6}{M}_{\odot}\,\mathrm{yr}^{-1}$] & [${M}_{\odot}$] & [Myr] \\
\midrule
$1.00\cdot10^{-6}$ & 3.75 & 1.85 & 1.49 & 1.75 & 0.51 & 3.23 & 1.49 & 1.50 & 1.76 & 0.51 \\
$1.58\cdot10^{-6}$ & 3.32 & 1.32 & 1.75 & 1.81 & 0.46 & 3.37 & 1.33 & 1.78 & 1.81 & 0.45 \\
$2.51\cdot10^{-6}$ & 3.41 & 1.10 & 2.19 & 1.84 & 0.39 & 3.45 & 1.09 & 2.25 & 1.85 & 0.37 \\
$3.98\cdot10^{-6}$ & 3.61 & 0.91 & 2.87 & 1.90 & 0.32 & 3.55 & 0.87 & 2.95 & 1.92 & 0.30 \\
$6.31\cdot10^{-6}$ & 3.85 & 0.72 & 3.96 & 1.98 & 0.25 & 3.89 & 0.69 & 4.16 & 1.98 & 0.24 \\
$1.00\cdot10^{-5}$ & 4.04 & 0.53 & 5.71 & 2.07 & 0.19 & 4.23 & 0.54 & 6.02 & 2.03 & 0.18 \\
$1.58\cdot10^{-5}$ & 4.19 & 0.38 & 8.43 & 2.15 & 0.14 & 3.33 & 0.27 & 8.62 & 2.26 & 0.13 \\
$2.51\cdot10^{-5}$ & 3.12 & 0.17 & 12.47 & 2.24 & 0.09 & 3.07 & 0.15 & 14.28 & 2.09 & 0.09 \\
$3.98\cdot10^{-5}$ & 3.04 & 0.11 & 19.08 & 2.30 & 0.06 & 3.04 & 0.11 & 18.86 & 2.40 & 0.06 \\
$6.31\cdot10^{-5}$ & 2.97 & 0.06 & 33.83 & 2.48 & 0.05 & 2.93 & 0.05 & 38.43 & 2.49 & 0.04 \\
$1.00\cdot10^{-4}$ & 2.96 & 0.04 & 44.97 & 2.49 & 0.02 & 2.91 & 0.04 & 49.74 & 2.54 & 0.03 \\
$3.98\cdot10^{-4}$ & 2.94 & 0.01 & 153.5 & 2.74 & 0.01 & 2.91 & 0.03 & 56.30 & 2.61 & 0.03 \\
\bottomrule
\end{tabular}
}
\tablefoot{
     Scenario described in Section \ref{cteperiod}.  Here are shown the mass  $M_{stop}$ and time $t_{stop}$ when the star reaches the stopping condition, the mean accretion rate of its evolution $\langle \dot{M} \rangle$ calculated according to Eq. \eqref{maverage}, the transition mass  $M_{\mathrm{tr}}$ and the transition time $t_{\mathrm{tr}}$ for different values of the period $P$ and the amplitude $A$. For these calculations, we assumed a metallicity of $Z=0.02$.
}
\end{table}

\subsection{Simulations for variations of the metallicity}
In Table~\ref{tab:summary_diferent_Z}, we summarize the results for the case of a constant period of the oscillations varying the metallicity.

\begin{table}[h]
\centering
\caption{Results for the oscillating mass accretion rate scenario (metallicity variations).}
\label{tab:summary_diferent_Z}
\resizebox{\hsize}{!}{
\begin{tabular}{l ccccc ccccc}
\toprule
\ \ \ \ \ \ Amplitude & 
\multicolumn{5}{c}{$Z=0.001$} & 
\multicolumn{5}{c}{$Z=0.01$} \\
\cmidrule(lr){2-6} \cmidrule(lr){7-11}
[${M}_{\odot}\,\mathrm{yr}^{-1}$] & 
$ M_{\mathrm{stop}}$ & $ t_{\mathrm{stop}}$& $\langle \dot{M} \rangle$ & $M_{\mathrm{tr}}$ & $t_{\mathrm{tr}}$ &
$ M_{\mathrm{stop}}$ & $ t_{\mathrm{stop}}$& $\langle \dot{M} \rangle$ & $M_{\mathrm{tr}}$ & $t_{\mathrm{tr}}$ \\
& 
[${M}_{\odot}$] & [Myr] & [$10^{-6}{M}_{\odot}\,\mathrm{yr}^{-1}$] & [${M}_{\odot}$] & [Myr] & 
[${M}_{\odot}$] & [Myr] & [$10^{-6}{M}_{\odot}\,\mathrm{yr}^{-1}$] & [${M}_{\odot}$] & [Myr] \\
\midrule
$1.00\cdot10^{-6}$ & 3.25 & 1.50 & $1.50 $ & 1.45 & 0.29 & 3.52 & 1.68 & $1.50$ & 1.68 & 0.45 \\
$1.58\cdot10^{-6}$ & 3.40 & 1.34 & $1.79$ & 1.49 & 0.26 & 3.68 & 1.50 & $1.79$ & 1.71 & 0.40 \\
$2.51\cdot10^{-6}$ & 3.57 & 1.15 & $2.23$ & 1.52 & 0.23 & 3.81 & 1.24 & $2.27$ & 1.72 & 0.33 \\
$3.98\cdot10^{-6}$ & 3.81 & 0.94 & $3.00$ & 1.52 & 0.19 & 4.14 & 1.04 & $3.01$ & 1.86 & 0.27 \\
$6.31\cdot10^{-6}$ & 4.20 & 0.75 & $4.25$ & 1.60 & 0.13 & 4.27 & 0.80 & $4.10$ & 1.88 & 0.22 \\
$1.00\cdot10^{-5}$ & 4.64 & 0.59 & $6.20$ & 1.70 & 0.10 & 4.72 & 0.60 & $6.16$ & 1.92 & 0.14 \\
$1.58\cdot10^{-5}$ & 5.11 & 0.44 & $9.31$ & 1.76 & 0.08 & 5.14 & 0.45 & $9.14$ & 2.12 & 0.11 \\
$2.51\cdot10^{-5}$ & 5.29 & 0.33 & $12.90$ & 1.80 & 0.07 & 5.30 & 0.33 & $12.90$ & 2.24 & 0.09 \\
$3.98\cdot10^{-5}$ & 6.35 & 0.25 & $21.70$ & 1.84 & 0.06 & 3.27 & 0.10 & $24.00$ & 2.28 & 0.07 \\
$6.31\cdot10^{-5}$ & 6.18 & 0.18 & $29.50$ & 1.87 & 0.05 & 3.23 & 0.07 & $30.30$ & 2.33 & 0.06 \\
$1.00\cdot10^{-4}$ & 4.61 & 0.08 & $48.30$ & 1.91 & 0.04 & 3.21 & 0.06 & $36.80$ & 2.36 & 0.05 \\
$1.00\cdot10^{-3}$ & 4.20 & 0.03 & $108.00$ & 2.01 & 0.02 & 3.17 & 0.03 & $84.4$ & 2.43 & 0.02 \\
$3.98\cdot10^{-3}$ & 4.15 & 0.02 & $173.00$ & 2.14 & 0.01 & 3.19 & 0.02 & $136.00$ & 2.60 & 0.01 \\
\bottomrule
\end{tabular}
}

\vspace{0.3cm}

\resizebox{\hsize}{!}{
\begin{tabular}{l ccccc ccccc}
\toprule
\ \ \ \ \ \ Amplitude  & 
\multicolumn{5}{c}{$Z=0.02$} & 
\multicolumn{5}{c}{$Z=0.03$} \\
\cmidrule(lr){2-6} \cmidrule(lr){7-11}
  [$\rm{M}_\odot\,\mathrm{yr}^{-1}$] & 
$ M_{\mathrm{stop}}$ & $ t_{\mathrm{stop}}$& $\langle \dot{M} \rangle$ & $M_{\mathrm{tr}}$ & $t_{\mathrm{tr}}$ &
$ M_{\mathrm{stop}}$ & $ t_{\mathrm{stop}}$& $\langle \dot{M} \rangle$ & $M_{\mathrm{tr}}$ & $t_{\mathrm{tr}}$ \\
& 
[${M}_{\odot}$] & [Myr] & [$10^{-6}{M}_{\odot}\,\mathrm{yr}^{-1}$] & [${M}_{\odot}$] & [Myr] & 
[${M}_{\odot}$] & [Myr] & [$10^{-6}{M}_{\odot}\,\mathrm{yr}^{-1}$] & [${M}_{\odot}$] & [Myr] \\
\midrule
$1.00\cdot10^{-6}$ & 3.31 & 1.54 & $1.51$ & 1.77 & 0.52 & 3.13 & 1.42 & $1.50$ & 1.80 & 0.54 \\
$1.58\cdot10^{-6}$ & 3.86 & 1.61 & $1.77$ & 1.83 & 0.46 & 3.23 & 1.25 & $1.79$ & 1.85 & 0.47 \\
$2.51\cdot10^{-6}$ & 3.45 & 1.08 & $2.27$ & 1.84 & 0.38 & 3.43 & 1.07 & $2.28$ & 1.93 & 0.41 \\
$3.98\cdot10^{-6}$ & 3.78 & 0.92 & $3.02$ & 1.94 & 0.31 & 3.74 & 0.91 & $3.03$ & 1.94 & 0.32 \\
$6.31\cdot10^{-6}$ & 4.81 & 0.90 & $4.22$ & 2.07 & 0.25 & 3.62 & 0.64 & $4.10$ & 2.12 & 0.26 \\
$7.23\cdot10^{-6}$ & 3.87 & 0.61 & $4.68$ & 2.07 & 0.23 & 3.85 & 0.60 & $4.73$ & 2.12 & 0.24 \\
$8.15\cdot10^{-6}$ & 4.12 & 0.60 & $5.21$ & 2.01 & 0.21 & 4.09 & 0.59 & $5.22$ & 2.10 & 0.23 \\
$1.00\cdot10^{-5}$ & 5.18 & 0.70 & $5.94$ & 1.98 & 0.18 & 3.82 & 0.47 & $5.99$ & 1.99 & 0.18 \\
$1.0725\cdot10^{-5}$ & 3.99 & 0.47 & $6.38$ & 2.00 & 0.15 & 3.97 & 0.45 & $6.53$ & 2.02 & 0.18 \\
$1.145\cdot10^{-5}$ & 4.14 & 0.45 & $6.92$ & 2.05 & 0.15 & 2.98 & 0.28 & $7.21$ & 2.05 & 0.15 \\
$1.29\cdot10^{-5}$ & 4.43 & 0.44 & $7.75$ & 2.13 & 0.13 & 2.97 & 0.25 & $7.80$ & 2.14 & 0.14 \\
$1.58\cdot10^{-5}$ & 3.25 & 0.25 & $9.14$ & 2.24 & 0.12 & 2.97 & 0.23 & $8.60$ & 2.27 & 0.12 \\
$2.51\cdot10^{-5}$ & 3.01 & 0.13 & $12.90$ & 2.37 & 0.09 & 2.81 & 0.11 & $16.20$ & 2.43 & 0.09 \\
$3.98\cdot10^{-5}$ & 2.95 & 0.09 & $22.60$ & 2.43 & 0.07 & 2.78 & 0.08 & $21.70$ & 2.49 & 0.08 \\
$6.31\cdot10^{-5}$ & 2.93 & 0.07 & $28.10$ & 2.47 & 0.06 & 2.76 & 0.07 & $26.70$ & 2.51 & 0.06 \\
$1.00\cdot10^{-4}$ & 2.92 & 0.06 & $33.90$ & 2.48 & 0.05 & 2.76 & 0.06 & $32.10$ & 2.58 & 0.05 \\
$1.00\cdot10^{-3}$ & 2.91 & 0.03 & $77.60$ & 2.63 & 0.02 & 2.76 & 0.02 & $77.60$ & 2.72 & 0.02 \\
$3.98\cdot10^{-3}$ & 2.94 & 0.02 & $118.00$ & 2.76 & 0.01 & 2.78 & 0.02 & $118.00$ & 2.78 & 0.02 \\
\bottomrule
\end{tabular}
}
\tablefoot{
     Scenario described in Section \ref{ctemetallicity}. For these calculations, we assumed an initial accretion rate of $\dot{M}_0 = 10^{-6}\ {M}_{\odot}\,\mathrm{yr}^{-1}$ and a period of $50,000$ years. $M_{\rm stop}$ and $t_{\rm stop}$ refer to the stopping condition.
}
\end{table}

\section{Assessment of the robustness of the criterion for the transition mass.}\label{robust}

\begin{figure}[H]
    \centering
    \includegraphics[width=0.75\hsize]{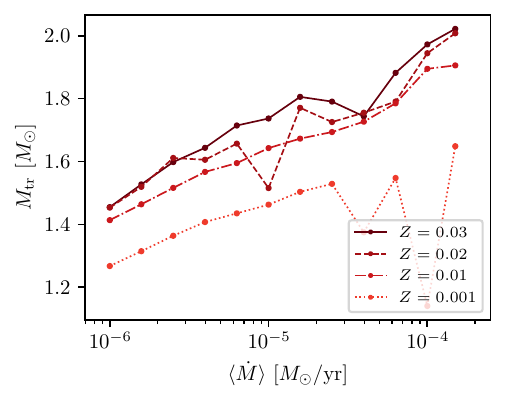}
    
    \vspace{0.4cm}
    
    \includegraphics[width=0.75\hsize]{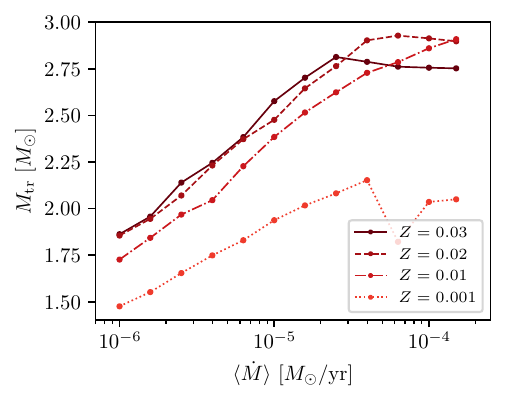}
    \caption{Variations of the criterion used in the research. \textit{Panel a (top):} The mass of the transition is defined as the mass when 35\% of the star's interior becomes radiative. \textit{Panel b (bottom):} The mass of the transition is defined as the mass when the star reaches a 60\% of radiativity in its interior.}
    \label{fig:dos_imagenes}
\end{figure}

Here we assess if and how much the mass of the transition from a convective to a radiative state depends on its exact definition. For this purpose, we vary the percentage of the star that must have become radiative for the definition of the mass of the transition. The corresponding results are given in Fig.~\ref{fig:dos_imagenes}, showing a case where we evaluate $M_{\mathrm{tr}}$ already when $35\%$ of the star is radiative, as well as a second case where we evaluate it at $60\%$. We notice here a small but systematic shift, as clearly the transition will occur somewhat later when a larger fraction of the star needs to become radiative. As can be seen from the supplementary Kippenhahn diagrams available on Zenodo, for the higher accretion rates above $\sim4\cdot10^{-6}\  M_\odot \ \mathrm{yr}^{-1}$ and mostly for metallicities $Z\geq0.01$, the outer parts of some of the stars never become radiative before the end of the calculation. In those cases it is possible that the magnetic field in the outer layers may always be dissipated by convection, even if in principle 3D magneto-hydrodynamic simulations would be in order to investigate this phenomenon.

\section{Dependence of the transition on the initial mass during the evolution}\label{apendixg}
\begin{figure}[h]
    \centering
    \includegraphics[width=0.9\hsize]{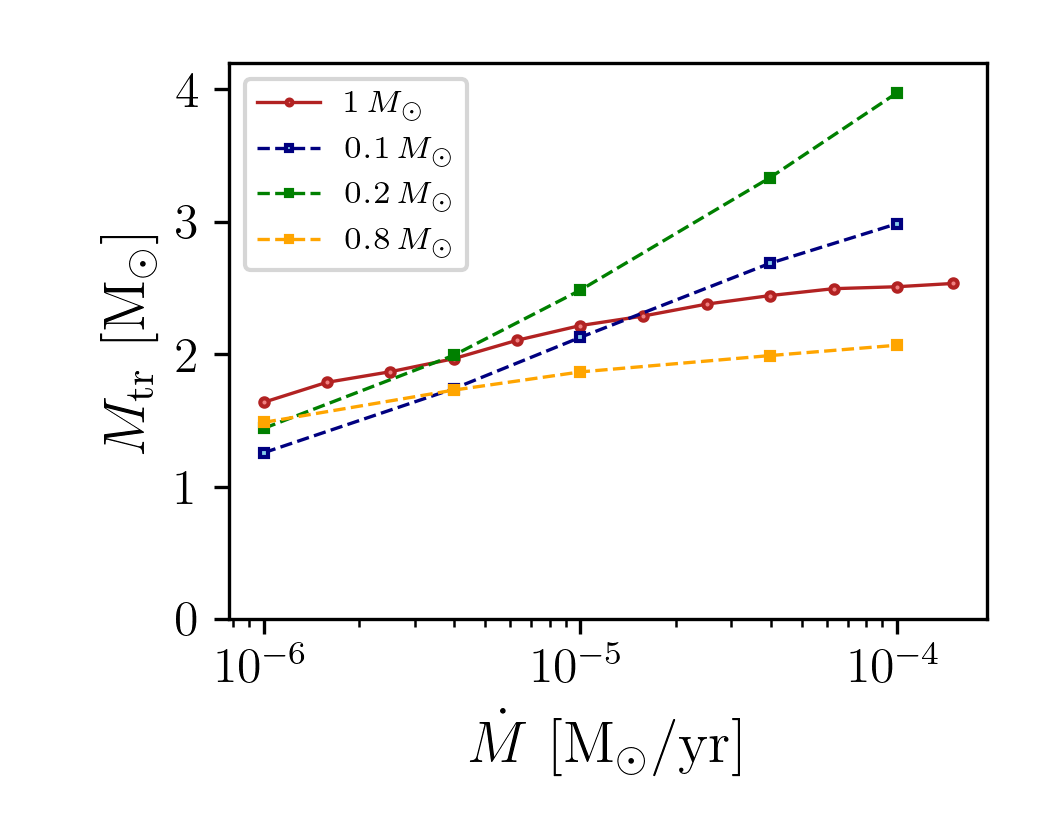}
    \caption{The critical mass for the transition as a function of the accretion rate.}
    \label{fig:lowerr}
\end{figure}

In Fig. \ref{fig:lowerr}, we investigate the dependency of the transition to a radiative state on the assumed initial mass of the protostar. In all cases, the protostar is initialized as a chemically homogeneous, fully convective $n=1.5$ polytrope. For accretion rates up to $\sim10^{-5}$~$M_\odot$~yr$^{-1}$, the dependence is rather {weak, with a spread in $M_{\rm tr}$ of $\lesssim 0.7\,M_\odot$}. For higher accretion rates, the spread {grows substantially, reaching $\approx 2\,M_\odot$ at $\dot M \approx 10^{-4}\,M_\odot\,\mathrm{yr}^{-1}$ ($M_{\rm tr} \approx 2.1$--$4.0\,M_\odot$)}. {This variation is} not systematic, but rather {depends} on the initial conditions. Thus, the mass at which the transition occurs should not be regarded as a unique fixed value; rather, some dispersion is expected around the point at which the transition occurs. {We note that these tests vary only the initial mass within the same polytropic construction, and therefore do not probe the thermal history inherited from self-consistent cold accretion.}

\section{Transition mass switching the deuterium abundance on and off.}\label{apendixh}

\begin{figure}[h]
    \centering
    \includegraphics[width=0.75\hsize]{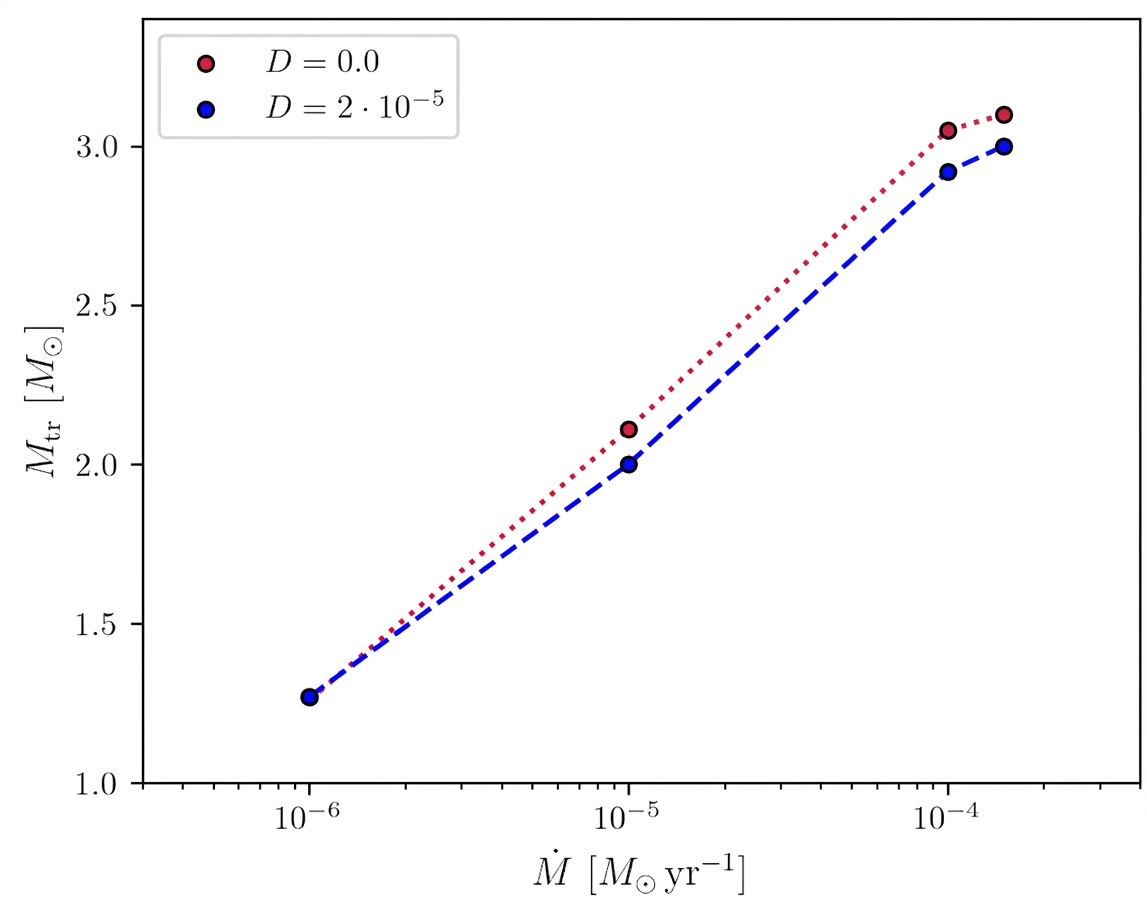}
    \caption{Mass of the transition as a function of the accretion {rate}. Red line indicates models without D (as in the main model grid), and the blue line indicates models with the specified D abundance.}
    \label{fig:D}
\end{figure}

In Fig. \ref{fig:D}, we simulate models with the deuterium burning switched on and off, with an initial mass of $0.15\,M_\odot$ and $Z=0.02$, to investigate its dependency. The mass where the transition occurs thus depends only weakly on the effects of deuterium burning. {The main model grid corresponds to the case without deuterium ($X_{\rm D}=0$, red curve), while the blue curve adopts $X_{\rm D}=2\cdot10^{-5}$. The difference in $M_{\rm tr}$ is $\lesssim 0.15\,M_\odot$ over the range of accretion rates explored.}
 
\end{appendix}

\end{document}